\documentclass[journal,twoside]{IEEEtran}

\usepackage{amsmath}
\usepackage{amssymb}
\usepackage{amsfonts}
\usepackage{amsthm}

\usepackage{calrsfs}
\usepackage[mathcal]{euscript}

\usepackage{cancel}
\usepackage{cite}
\usepackage{epsf}

\usepackage{subfigure}
\usepackage{multicol}
\usepackage[usenames]{color}

\newcommand{\RR}{\mathbb{R}}

\DeclareMathOperator{\prob}{\mathbb{P}}
\DeclareMathOperator{\ex}{\mathbb{E}}

\DeclareMathOperator{\supp}{supp}

\newcommand{\dd}{\,d}
\newcommand{\eps}{\varepsilon}

\newcommand{\outage}{P_{\text{out}}}
\newcommand{\energy}{\mathcal{E}}
\newcommand{\lagrange}{\mathcal{L}}
\newcommand{\hilbert}{\mathcal{T}}
\newcommand{\bigoh}{\mathcal{O}}
\newcommand{\tame}{\Omega}
\newcommand{\domain}{\mathcal{X}}
\newcommand{\DD}{\,\mathcal{D}}
\newcommand{\pf}{\mathcal{Z}}

\definecolor{blue}{rgb}{0,0,1}
\definecolor{olivegreen}{rgb}{0,0.5,.2}
\definecolor{orange}{rgb}{1,0.4,0}

\theoremstyle{plain}
\newtheorem{theorem}{Theorem}
\newtheorem{proposition}[theorem]{Proposition}

\newtheorem{lemma}[theorem]{Lemma}

\theoremstyle{definition}
\newtheorem{conjecture}[theorem]{Conjecture}
\newtheorem{definition}[theorem]{Definition}

\theoremstyle{remark}
\newtheorem*{remark*}{Remark}
\newtheorem{remark}{Remark}

\numberwithin{remark}{theorem}

\newfont{\bbb}{msbm10 scaled 1100}
\newcommand{\lambdav}{\hbox{\boldmath$\lambda$}}
\newcommand{\Dc}{{\cal D}}
\newcommand{\Ec}{{\cal E}}

\def\bH{{\bf H}}

\def\bI{{\bf I}}
\newfont{\bb}{msbm10 scaled 1100}

\begin{document}
\bstctlcite{BSTcontrol}

\title{Living at the Edge: A Large Deviations Approach to the Outage MIMO Capacity} %

\author{Pavlos Kazakopoulos, Panayotis Mertikopoulos, Aris L. Moustakas and Giuseppe Caire
\thanks{P. Kazakopoulos (pkazakop@phys.uoa.gr), P. Mertikopoulos (pmertik@phys.uoa.gr) and A. L. Moustakas (arislm@phys.uoa.gr) are with the Physics Dept., Athens Univ., 157 84  Athens, Greece. G. Caire (caire@usc.edu) is with the EE - Systems Dept., Univ. Southern California Los Angeles, CA 90007, USA. Part of this paper was presented in the 2009 IEEE Information Theory Workshop (ITW '09) in Volos, Greece. This research was supported in part by Greek GSRT "Kapodistrias" project No. 70/3/8831.}}


\maketitle

\begin{abstract}
A large deviations approach is introduced, which calculates the probability density and outage probability of the MIMO mutual information, and is valid for large antenna numbers $N$. In contrast to previous asymptotic methods that only focused on the distribution close to its {\it most probable} value, this methodology obtains  the {\it full} distribution, including its non-Gaussian tails. The resulting distribution interpolates between the Gaussian approximation for rates $R$ close its mean and the asymptotic distribution for large signal to noise ratios $\rho$ \cite{Zheng2003_DiversityMultiplexing}. For large enough $N$, this method provides the outage probability over the whole $(R, \rho)$ parameter space. The presented analytic results agree very well with numerical simulations over a wide range of outage probabilities, even for small $N$. In addition, the outage probability thus obtained is more robust over a wide range of $\rho$ and $R$ than either the Gaussian or the large-$\rho$ approximations, providing an attractive alternative in calculating the probability density of the MIMO mutual information. Interestingly, this method also yields the eigenvalue density constrained in the subset where the mutual information is fixed to $R$ for given $\rho$. Quite remarkably, this eigenvalue density has the form of the Mar\v{c}enko-Pastur distribution with square-root singularities.
\end{abstract}


\section{Introduction}
\label{Introduction}

Considerable interest has arisen from the initial prediction\cite{Foschini1998_BLAST1, Telatar1995_BLAST1} that the use of multiple antennas in transmitting and receiving signals can lead to substantial gains in information throughput. To analyze the theoretical limits of such a MIMO (Multiple Input Multiple Output) system, it has been convenient to focus on the case of i.i.d. Gaussian noise and input. For the MIMO channel model
\begin{equation}\label{eq:MIMO_ch_model}
\mathbf{y= Hx + z}
\end{equation}
with coherent detection and no channel state information at the transmitter \cite{Foschini1998_BLAST1, Telatar1995_BLAST1}, the mutual information $I_N$ for a given value of the channel matrix $\bH$ takes the familiar form:
\begin{equation}\label{eq:logdet_def}
I_N=\log\det\left(\bI + \rho\bH^\dagger\bH\right).
\end{equation}
where ``$\log$'' signifies the natural logarithm, $\rho$ is the signal to noise ratio and  $\bH$ is the  $M\times N$ channel matrix whose elements are independent ${\cal CN}(0,1/N)$ random variables. This corresponds to the case of $N$ transmitting and $M$ receiving antennas, which is captured by the ratio $\beta=M/N$. Without loss of generality we assume that $\beta\geq 1$; otherwise, if $\beta<1$, we may simply replace $\rho$ with $\rho_{\text{new}}=\rho\beta$ in (\ref{eq:logdet_def}) and interchange the roles of $M$ and $N$.

If the channel matrix $\bH$ varies in time according to a stationary ergodic process, and coding spans an arbitrarily large number of fading states, then the ``ergodic'' channel capacity is given by the mutual information expected value $\ex\left[I_N\right]$ \cite{Telatar1995_BLAST1}. Initially, this quantity was calculated asymptotically for large $N$, with $\beta$ remaining fixed and finite. In particular, in this case, $\bH$ can be viewed as a large random matrix. Then, by applying ideas and methods from the theory of random matrices, it was shown in \cite{Rapajic2000_InfoCapacityOfARandomSignatureMIMOChannel} that the value of the mutual information per antenna $I_N(\rho,\bH)/N$ ``freezes'' to a deterministic value in the large $N$ limit,
the so-called {\it ergodic average} $r_{\text{erg}}(\rho)$. Underlying this result is the fact that the very eigenvalue distribution of $\bH^\dagger\bH$ freezes to the celebrated Mar\v{c}enko-Pastur distribution:
\begin{equation}
\label{eq:MP_def}
p(x) = \frac{\sqrt{(b-x)(x-a)}}{2\pi x}
\end{equation}
where $a, b= (\sqrt{\beta}\pm 1)^2$ are the end-points of its support. Even though later the closed form solution of $\ex[I_N]$ for general $M$, $N$ was found\cite{Wang2002_OutageMutualInfoOfSTMIMOChannels}, the asymptotic form of $r_{erg}(\rho)$ was particularly popular due to its simplicity and accuracy, even for small number of antennas.

Another more relevant regime is when the channel matrix is random, but varies in time much more slowly than the typical coding delay. In this case (usually referred to as the ``quasi-static'' fading channel) $\bH$ can be considered as a random constant and the mutual information $I_N(\bH)$ is  a random variable. In this regime, the relevant performance metric is the ``rate versus outage probability'' tradeoff \cite{Biglieri1998_FadingChannels}, captured by the cumulative distribution function of $I_N(\bH)$. Various approaches \cite{Moustakas2003_MIMO1, Hochwald2002_MultiAntennaChannelHardening, Hachem2006_GaussianCapacityKroneckerProduct, Taricco2006_MIMOCorrelatedCapacity, Taricco2008_MIMOCorrelatedCapacity} have shown that the mutual information $I_N(\bH)$ becomes asymptotically Gaussian for large $N$, with mean equal to the ergodic capacity $R_{\text{erg}} = N r_{\text{erg}}(\rho)$ and a variance of order $\bigoh (1)$ in $N$. This Gaussian variability of the mutual information is due to the fluctuations of the eigenvalues of the matrix around the most probable distribution described by the Mar\v{c}enko-Pastur law. Since this Gaussian approximation is essentially a variation of the central limit theorem, it only applies within a small number of standard deviations away from the mean $R_{\text{erg}}$. As a result, this approximation fails to capture the tails of the distribution, e.g. the probability of the mutual information $I_N$ falling below half its ergodic value $R_{\text{erg}}/2$, because this event only occurs $\bigoh(N)$ standard deviations away from the mean.

Nevertheless, the tails of the distributions of the mutual information are important, because they correspond to regions with low outage probability, where one would want to operate a MIMO system. This is particularly important when, for large $\rho$, the slopes of the outage curves are large. The interplay between low outage and multiplexing gain was exemplified in the seminal paper \cite{Zheng2003_DiversityMultiplexing} where the authors analyzed the asymptotics of the distribution of the mutual information in the limit of large $\rho$ (keeping $R/\log\rho$ fixed). They found that the asymptotic form of the logarithm of the outage probability  of the mutual information $\outage(R) \equiv \prob(I_{N}(\bH)\leq R)$   is a piecewise linear function of $R/\log\rho$, interpolating between the discrete set of values:
\begin{equation}\label{eq:DMT_def}
\log \outage (R_n) \sim - \log\rho\left(\frac{R_n}{\log\rho} - M\right)\left(\frac{R_n}{\log\rho} - N\right)
\end{equation}
where $R_n = n \log\rho$ for integer $n\leq N\leq M$. When, in addition to $\rho$, $N$ is also large, $\log \outage(R)$ in (\ref{eq:DMT_def}) becomes (to leading order) a continuous function of $R/N$. It should be pointed out that this approach generalizes the large $N$ asymptotics discussed above, since it provides insight in the distribution of the mutual information quite far from its peak, which for large $\rho$ (and large $N$) is situated at $I_N\approx N\log\rho$.  More recently, in \cite{Azarian2007_finite_rate_DMT} the authors recast the DMT problem providing a formula to calculate $\log P_{out}$ as a function of $R$ when $R$ lies in each linear subsegment of (\ref{eq:DMT_def}). Nevertheless both approaches \cite{Zheng2003_DiversityMultiplexing, Azarian2007_finite_rate_DMT} do not provide the offset to the leading, $O(\log\rho)$ behavior of (\ref{eq:DMT_def}). As a result, these approaches, while quite intuitive fail, often by a large margin, to provide an acceptable quantitative estimate of $P_{out}$ unless $\log\rho$ is extremely large.

In the meantime, all variants \cite{Moustakas2003_MIMO1, Hochwald2002_MultiAntennaChannelHardening, Hachem2006_GaussianCapacityKroneckerProduct} of the large $N$ Gaussian approximation of the mutual information fail for large $\rho$. Specifically, they all predict that the outage probability is given asymptotically by:
\begin{equation}
\label{eq:largeN_large_rho}
\log \outage (R) \sim  \frac{(\log\rho)^2}{2\log\left(1-\beta^{-1}\right)}\left(\frac{R}{\log\rho} - N\right)^2
\end{equation}
where $\beta=M/N>1$, an expression which is in striking disagreement with (\ref{eq:DMT_def}). Even though for $\beta=1$ the asymptotic form of (\ref{eq:DMT_def}) is recovered within the Gaussian approximation\cite{Moustakas2003_MIMO1, Hachem2006_GaussianCapacityKroneckerProduct}, the discrepancy for $\beta\neq 1$ indicates that the limits $N\rightarrow\infty$ and $\rho\rightarrow\infty$ cannot be na\"ively interchanged. In the Gaussian approximation,  one focuses on the most probable eigenvalue distribution, which converges vaguely to the Mar\v{c}enko-Pastur distribution (\ref{eq:MP_def}). However, as can be seen in (\ref{eq:MP_def}), this distribution (almost surely) produces no eigenvalues of $\bH^\dagger\bH$ close to zero when $\beta>1$. Nevertheless, the analysis for large $\rho$ focuses at the regime where the eigenvalues are of order  $O(\rho^{-1})$. As a result, it is not surprising that the large-$N$ Gaussian approximation of the mutual information distribution misses the correct behavior.

In summary, we have two methods, the large-$N$, fixed-$\rho$ Gaussian approximation on the one hand and the large-$\rho$, fixed-$N$ limit on the other, both having their own regions of validity, and both failing to produce quantitative results for the outage probability outside their respective regions. Thus, one still needs an approach that correctly describes the outage behavior of the mutual information distribution for arbitrary $\rho$ and $R$.

In this paper, we introduce a large deviations approach to calculate the full asymptotic distribution of $R$. It is formally valid for large $N$, but works over the whole range of values of $R$ and $\rho$. This method bridges the two regions of small/intermediate and large signal to noise ratios within a single framework and, in effect, it amounts to calculating the rate function of the logarithm of the average moment generating function of the mutual information. Our approach was first introduced in the context of random matrix theory by Dyson \cite{Dyson1962_DysonGas} and has been more recently applied in a variety of problems \cite{Majumdar2006_LesHouches, Vivo2007_LargeDeviationsWishart, Vivo2008_DistributionsConductanceShotNoise, Nadal2009_NonIntersectingBrowianInterfaces}. It is quite intuitive because it interprets the eigenvalues of $\bH^\dagger\bH$ as point charges on a line repelling each other logarithmically. This is the first time this approach has been applied in information theory and communications. As a byproduct of this approach, we obtain the most probable eigenvalue distribution constrained on the subset of channel matrices $\bH^\dagger\bH$ that have fixed total rate $R$ and signal to noise ratio $\rho$. This is a generalized Mar\v{c}enko-Pastur distribution that gives the constrained eigenvalue distribution for values of $R$ even far from its ergodic value. It is worth pointing out that many of the results presented here could be set on a more formal mathematical footing using tools developed in \cite{Johansson1998_2ndOrderRMTFluctuations}. However, we will follow the less formal but more intuitive approach developed by Dyson.
%
%

This generalized Mar\v{c}enko-Pastur distribution can also be seen as the inverse of the so-called {\it Shannon transform} \cite{Tulino2004_RMTInfoTheoryReview} in the following sense: while the Shannon transform produces the value of normalized mutual information $I_N/N$ as a functional of the asymptotic eigenvalue distribution of $\bH^\dagger\bH$ (the Mar\v{c}enko-Pastur distribution), the generalized Mar\v{c}enko-Pastur distribution introduced here boils down to the asymptotic eigenvalue distribution of $\bH^\dagger\bH$ for a given value of the mutual information $R=Nr$, i.e., when $\bH^\dagger\bH$ is constrained on the subset defined by $r= I_N(\bH)/N$.

\subsection{Outline}
\label{sec:Outline}

In the next section we will introduce the necessary mathematical methodology. In particular, Section \ref{sec:map_coulomb_gas} describes the mapping of the joint probability distribution of eigenvalues of the Wishart matrix to a Coulomb gas of charges with a continuous density (discussed in more detail in Appendix \ref{app:Coulomb_gas}) and the large-deviations analysis of the problem. Next, section \ref{sec:integral_equation} deals with the solution of the resulting integral equation that produces the most-likely eigenvalue distribution at the tails of the full distribution.

If one is not particularly interested in the details of our derivation, Section \ref{sec:methodology} may be skipped in favor of section \ref{sec:results} where we present our main results. Specifically, in Section \ref{sec:MP_Distribution} we rederive the Mar\v{c}enko-Pastur distribution (that is, the most likely distribution without the mutual information constraint) to highlight the efficacy of our method. Subsequently, Sections \ref{sec:beta>1} and \ref{sec:beta=1} contain our results for the cases $\beta>1$ and $\beta=1$ respectively, while in Section \ref{sec:OutageProbability} we show how to calculate the outage probability directly by means of the results of the previous sections. In Section \ref{sec:limiting_cases}  we analytically obtain previous results as limiting cases of this method, and also examine a number of different limiting cases. In Section \ref{sec:numerical simulations} we provide numerical comparisons of our method to other approximations previously outlined and to Monte Carlo simulations.

The proofs of the properties of {\it tame} distributions (introduced in section \ref{sec:map_coulomb_gas}) are given in appendix \ref{app:tameness} and we discuss Dyson's original construction of the Coulomb gas model in appendix \ref{app:Coulomb_gas}. Appendices \ref{app:solution} and \ref{app:uniqueness_ab} have been reserved for the exposition of some technical issues that cropped up during our calculations. Finally, Appendix \ref{app:1/Ncorrections} discusses higher order $O(1/N)$ corrections to our model and comparisons with Monte Carlo simulations.

\section{Methodology}
\label{sec:methodology}

Our approach can roughly be divided in two main parts. First, in section \ref{sec:map_coulomb_gas} we reduce the original problem of finding the probability distribution of the mutual information to harvesting the minimum energy of a gas of charged particles (among other things we show here that the minimum energy configuration is unique). Then, in section \ref{sec:integral_equation}, we will solve the integral equation that comes up and actually obtain the minimum energy configuration of the charges.


\subsection{Mapping the Problem to a Coulomb Gas}
\label{sec:map_coulomb_gas}

We begin by establishing the mathematical methodology, treading on the elegant footsteps of \cite{Vivo2007_LargeDeviationsWishart, Dean2008_ExtremeValueStatisticsEigsGaussianRMT}. Our overall aim will be to calculate the probability distribution of the mutual information (\ref{eq:logdet_def}), which can be written in terms of the eigenvalues $\lambda_{k}$ of the Wishart matrix $\bH^\dagger\bH$ as:
\begin{eqnarray}
\label{eq:MI_def}
I_N(\lambdav) &=& \sum_{k=1}^N \log\left(1+\rho\lambda_k\right)
\end{eqnarray}

Note that the aforementioned probability distribution of the mutual information thus depends on the joint probability distribution function of the eigenvalues $\lambda_1\ldots \lambda_{N}$ of $\bH^\dag\bH$. In its turn, this distribution takes the well-known form:
\begin{eqnarray}
\label{eq:P_lambda_k}
P_{{\lambdav}}(\lambda_{1}\ldots\lambda_{N}) &=& A_N
\Delta(\lambdav)^2 \prod_{k=1}^N \lambda_k^{M-N} e^{- N \lambda_k} \\
&=& A_N e^{-N^2 E\left(\lambdav\right)}
\label{eq:P_lambda_k_exp}
\end{eqnarray}
where $A_N$ is a normalization constant and $\Delta(\lambdav) = \prod_{i>j} (\lambda_i-\lambda_j)$ is the Vandermonde  determinant of the eigenvalues $\lambda_k$.
The exponent $E(\lambdav)$ is an energy function of the eigenvalues $\{\lambda_i\}$  that will become very useful later:
\begin{eqnarray}\label{eq:E_lambda}
E(\lambdav)&=& \frac{1}{N} \sum_k \left(\lambda_k - (\beta-1) \log\lambda_k\right) \\ \nonumber
&+& \frac{2}{N^2} \sum_{j>k} \log\left|\lambda_j-\lambda_k\right|
\end{eqnarray}
Note that the normalization we have chosen is such that $E(\lambdav)$ corresponds roughly to the energy per eigenvalue.

The cumulative probability distribution (CDF) of the normalized mutual information $I_N/N$ can then be written as a ratio of two volumes in $\lambdav$-space:
\begin{eqnarray}
\label{eq:CDF_vol_ratio}
F_{N}(r)&=& \prob(I_N/N\leq r)
 		= \frac{V_r}{V_{\text{tot}}} \\ \nonumber
		&=&\int P_{\lambdav}(\lambdav)\, \Theta(r-I_N/N)\dd\lambdav
\end{eqnarray}
where $I_N$ is given by (\ref{eq:MI_def}), {$\Theta(x)$ is the Heaviside step function ($\Theta(x)=1$ if $x>0$ and $\Theta(x)=0$ if $x<0$) and the integrals are taken with respect to the ordinary $N$-dimensional Lebesgue measure $d\lambdav =\prod_i d\lambda_i$}. The above CDF is by definition the outage probability, i.e. the probability that the normalized mutual information falls below $r$. Its corresponding probability density (PDF) can be obtained from (\ref{eq:CDF_vol_ratio}) by taking the derivative with respect to $r$ \cite{Papoulis_book}:
\begin{equation}
\label{eq:P_R_vol_ratio}
 P_{N}(r)		= F_{N}'(r)
 			=\int P_{\lambdav}(\lambdav) \, \delta(r-I_N/N) \dd\lambdav
\end{equation}
where we have used the fact that the (distributional) derivative of the step function is the Dirac $\delta$-function: $\Theta'(x) = \delta(x)$.

Our primary goal will be to use (\ref{eq:P_R_vol_ratio}) in order to obtain an analytic expression for the probability distribution function of the mutual information $I_{N}$. However, in general there is no standard way to evaluate integrals like $V_r$ (except for some special cases \cite{Simon2002_TIMO1}). Nevertheless, in the large-$N$ limit it is possible to analyze such integrals in a systematic way. This so-called Coulomb-gas approach \cite{Mehta_book} is based on the intuitive idea to interpret the eigenvalues $\lambda$ as the positions of $N$ positive unit charges located on a line, a picture first proposed by Dyson \cite{Dyson1962_DysonGas}. Within this interpretation, the last term in the exponent $E(\lambdav)$ in (\ref{eq:E_lambda}) corresponds to the logarithmic repulsion energy, while the first term is the potential due to a constant field and the second term is the repulsion of a point charge located at the origin.\footnote{Note that these are simply the potentials that one obtains in classical two-dimensional electrostatics.}

Now, it is instructive to look at the form of $E(\lambdav)$ to get an intuitive understanding of the minimum energy configuration of $\lambdav$ in the absence of the constraint $I_N/N=r$. As discussed above, the first two terms in $E(\lambdav)$ correspond to the external forces acting on the charges, while the last term represents the repulsion between charges. In the absence of the charge repulsion the minimum energy configuration will correspond to all charges settling at the minimum of the external potential, i.e. $\lambda_k=\beta-1$ for all $k=1,\ldots,N$. However, the repulsion between charges will make them move away from that point but still, from simple electrostatics considerations, the external forces will not allow this repulsion to carry charges too far away from the minimum. As a result, we expect that at the minimum of $E(\lambdav)$ all charges will be concentrated in the neighborhood of $\beta-1$. As the number of charges increases, it will make sense, at least for configurations with energy $E(\lambdav)$ close to the minimum, to expect that the charge distribution will be approximately a {\it continuous} distribution. As a result, all sums over $\lambdav$ in $E(\lambdav)$ may be replaced by integrals, and we expect that this will also be true in the presence of constraints as in (\ref{eq:P_R_vol_ratio}).

To make this continuum limit more precise, one begins by conditioning the probability law $\prob$ of the eigenvalues of the Wishart matrix $\bH^{\dagger}\bH$ on the set $J_{r}=\{\lambda:I_{N}(\lambda)/N=r\}$, i.e. by considering the conditional probability law $\prob(\cdot|I_{N}/N=r)$ and the corresponding PDF. As $N\to\infty$, large deviations theory suggests that this density function will be sharply concentrated around its most probable value, i.e. the minimum of the energy functional (\ref{eq:E_lambda}). Then, according to Dyson, this minimum can be asymptotically recovered by looking at the minimum of the {\it continuous} version of (\ref{eq:E_lambda}):
\begin{conjecture}[Coulomb Gas Assumption]
As $N\to\infty$, the empirical distribution of charges/eigenvalues under the rate constraint $I_{N}/N = r$ converges vaguely to an absolutely continuous density $p(x)$ which minimizes the continuous energy functional:
\begin{eqnarray}
\label{eq:energy}
\energy[p]		&=& \int \!x  p(x) \dd x- (\beta-1)\int\! p(x) \log x \dd x \\ \nonumber
			&-&\iint p(x) p(y) \!\log|x-y| \dd x dy
\end{eqnarray}
over the space of densities which satisfy the constraint $\int_{0}^{\infty}p(x)\log(1+\rho x) \dd x=r$.
In other words, as $N\to\infty$, the total charge in any interval $I\subseteq\RR$ will be given by:
\begin{equation}
\sigma(I) = \int_{I} p(x) dx,
\end{equation}
with $p$ as above.
\end{conjecture}

This assumption is essentially identical to the one in Mehta's book \cite{Mehta_book} and has been extensively employed in the literature \cite{Vivo2007_LargeDeviationsWishart, Dyson1962_DysonGas, Majumdar2006_LesHouches}. Unfortunately, despite its simple and intuitive nature, this assumption has resisted most attempts at a rigorous proof, thereby giving birth to different approaches, such as the one in \cite{Johansson1998_2ndOrderRMTFluctuations}. Nevertheless, the results obtained there are in agreement with the ones obtained with the help of the Coulomb Gas assumption and, hence, we feel that our posit here is rather mild (see also appendix \ref{app:Coulomb_gas} for a more detailed discussion).

At any rate, to make proper use of the energy functional $\energy$ (\ref{eq:energy}) we must first make sure that it remains finite over a reasonably large class of densities $p(x)$. This leads us to the concept of ``tameness'':
\begin{definition}
\label{dfn:tame}
An {integrable} function $p:\RR_{+}\to\RR$ will be called $\eps$-{\it tame} when:
\begin{itemize}
\item[(i)] the ``absolute mean'' of $p$ is finite:
\begin{equation}
\label{eq:center}
\int_{0}^{\infty} x |p(x)| \dd x<\infty;
\end{equation}
\item[(ii)] there exists some $\eps>0$ such that $p$ is $L^{1+\eps}$-integrable, i.e.
\begin{equation}\label{eq:1+e_integrable_def}
\int_{0}^{\infty}|p(x)|^{1+\eps} \dd x < \infty.
\end{equation}
\end{itemize}
\end{definition}

\begin{remark}
The phrasing of condition (i) simply reflects our interest in tame functions $p\equiv p_{X}$ that are probability densities of random variables $X$ with values in $\RR_{+}$. In that case, condition (i) simply states that $X$ has finite mean:
\begin{equation}
\tag{\ref{eq:center}'}
\ex\left[X\right]= \int_0^\infty x p(x)\dd x <\infty.
\end{equation}
\end{remark}

\begin{remark}
Condition (ii) will be crucial to our analysis. At first, it might appear as a mere technical necessity (see e.g. section \ref{sec:integral_equation} and appendix \ref{app:solution}) but, in fact, it has a very deep physical interpretation: a probability density with finite mean might still fail to have finite energy, making it inadmissible on physical grounds. Condition (ii) ensures that $\energy[p]$ will be finite (see lemma \ref{lem:tame} below).
\end{remark}

\begin{remark}
When it is not necessary to make explicit mention of the exponent $\eps$, we will simply say that $p$ is {\it tame}. Similarly, an absolutely continuous (signed) measure $\sigma$ on $\RR_{+}$ will be called {\it tame} when its Lebesgue derivative $p(x) = \frac{d\sigma(x)}{dx}$ is tame. Given this equivalence between continuous measures and Lebesgue derivatives, we will use the two terms interchangeably.
\end{remark}

Going back to the energy functional $\energy$ of (\ref{eq:energy}), we can see that condition (i) guarantees that the first term in (\ref{eq:energy}) is finite, while (ii) bounds the second and third terms. This is captured in the following:
\begin{lemma}[Finiteness and Continuity of $\energy$]
\label{lem:tame}
Let $\tame$ be the space of tame functions on $\RR_{+}$ and let $\energy$ be defined as in (\ref{eq:energy}). Then, $\energy[p]<\infty$ for all $p\in\tame$ and the restriction of $\energy$ to any subspace of $L^{1+\eps}$-integrable functions with finite mean is continuous (in the $L^{1+\eps}$ norm). In other words,  tame densities have finite energy and tame variations in density induce small variations in energy.
\end{lemma}

We prove this lemma in Appendix \ref{app:tameness} where we also give some background information on the $L^{r}$ norms. For now, it will be more useful to express the probability density $P_{N}(r)$ as the ratio:
\begin{equation}
P_{N}(r) = \frac{\pf_{r}}{\pf}
\end{equation}
where, in accordance with (\ref{eq:P_lambda_k_exp}), (\ref{eq:P_R_vol_ratio}) and (\ref{eq:energy}), $\pf_{r}$ and $\pf$ are the (un-normalized) {\it partition functions}:\footnote{It is worth pointing out that the correction to the term $N^2\energy[p]$ in the exponent is $\bigoh(1)$ (see appendix \ref{app:Coulomb_gas} for more details). Also a nice analysis of the mapping from the $\lambdav$ integrals to path integrals over $p$ can also be found in \cite{Dean2008_ExtremeValueStatisticsEigsGaussianRMT}.}
\begin{eqnarray}\label{partition_functions_def}
\pf_{r} = \int_{\domain_{r}}\!\!\!\DD \!p \,\, e^{-N^{2}\energy[p]}\\
\pf = \int_{\domain}\!\!\!\DD \!p \, \, e^{-N^{2}\energy[p]}
\end{eqnarray}
and $\DD p$ denotes the path-integral measure over the domains of tame densities $\domain,\domain_{r}\subseteq\tame$:
\begin{eqnarray}
\domain = \left\{p\in\tame: p\geq 0 \text{ and } \int\! p(x)\dd x =1\right\}\\
\domain_{r} = \left\{p\in\domain:  \int\! p(x) \log(1+\rho x) \dd x = r\right\}.
\end{eqnarray}

Of course, from a mathematical point of view, constructing a measure $\DD p$ over the infinite-dimensional space of functions is an intricate process which is far from trivial. Path integrals were first introduced by R. Feynman \cite{Feynman1965_QM_PathIntegrals} in physics and have been used there extensively over the last 70 years.
We prefer not to introduce them formally but, rather, to follow a more intuitive approach instead, in Appendix \ref{app:Coulomb_gas}.

With all these considerations taken into account, we may take the large $N$ limit and write:
\begin{equation}
\lim_{N\to\infty} \frac{1}{N^{2}} \log P_{N}(r) =
\lim_{N\to\infty} \frac{1}{N^{2}} \left(\log\pf_{r} - \log\pf\right)
\end{equation}
and, by invoking Varadhan's lemma\cite{Dembo_book_LargeDeviationsTechniques}, we obtain:
\begin{equation}
\lim_{N\to\infty} \frac{1}{N^{2}} \log P_{N}(r) = \energy_0 - \energy_1(r)
\end{equation}
or, equivalently:
\begin{equation}
P_{N}(r) \sim e^{-N^{2} \left(\energy_{1}(r) - \energy_{0}\right)}
\label{eq:P_N(r)}
\end{equation}
where
\begin{eqnarray}\label{eq:E0=infE}
\energy_{0} &=& \inf_{p\in\domain} \energy[p]
\\ \label{eq:E1=infE}
\energy_{1}(r) &=& \inf_{p\in\domain_{r}} \energy[p]
\end{eqnarray}
In other words, we have reduced the problem of determining the asymptotic behavior of $P_{N}(r)$ to finding the minimum of the convex functional $\energy$ over the two convex domains $\domain$ and $\domain_{r}$. To that end, we have:
\begin{lemma}[Convexity of $\energy$]
\label{lem:convex}
Let $\domain\subseteq\Omega$ be the set of tame probability measures: $\domain = \left\{p\in\Omega: p\geq 0 \text{ and } \int\!p(x) \dd x = 1\right\}$. Then, $\domain$ is a convex subset of the topological vector space $\tame$ and $\energy$ is (strictly) convex on $\domain$.
\end{lemma}
Again, we will postpone the proof of this lemma until appendix \ref{app:tameness}. However, an immediate corollary is that there exists a unique charge density $p$ which minimizes (\ref{eq:E0=infE}) and (\ref{eq:E1=infE}). To find this unique solution - and the corresponding (global) minima $\energy_{0}, \energy_{1}(r)$ - it turns out to be more convenient to work over the whole space of tame measures $\tame$ and introduce Lagrange multipliers for the two domains $\domain$ and $\domain_{r}$. This leads to the Lagrangian functions:
\begin{eqnarray}
\label{eq:L0}
\lagrange_{0}[p,\nu,c]	&=& \energy[p] - c\left(\int_0^\infty\!\!\! p(x) \dd x -1\right) \nonumber\\
					&-& \int_0^\infty \!\!\!\nu(x)p(x) \dd x
\end{eqnarray}
\begin{eqnarray}
\label{eq:L1}
\lagrange_{1}[p,\nu,c,k]	&=& \lagrange_0[p,\nu,c]\nonumber\\
					&-& k\left(\int_0^\infty\!\!\! p(x)\log(1+\rho x) \dd x-r\right)
\end{eqnarray}
from which we obtain $\energy_{0}$ and $\energy_{1}(r)$ by maximizing over the dual parameters $\nu$ (non-negativity constraint), $c$ (normalization constraint) and $k$ (mutual information constraint):
\begin{eqnarray}
\label{eq:minimum_0}
\energy_0		&=& \sup_{\nu\geq 0; \, c} \inf_p \lagrange_0[p,\nu,c]\\
\energy_1(r)	&=& \sup_{\nu\geq 0; \, c,k} \inf_p \lagrange_1[p,\nu,c,k]
\label{eq:minimum_1}
\end{eqnarray}
The convexity of $\lagrange_0$, $\lagrange_1$ over $p$ ensures that it suffices to find a local minimum $p(x)$ for the corresponding Lagrangian $\lagrange$, for fixed $\nu$, $c$, $k$. Then, any value of $k$, $c$ that satisfies the constraints of $p$ will be unique \cite{Boyd_book}. It is also worth pointing out that the only difference between $\energy_0$ and $\energy_1$ above is that the former can be seen as the maximum over $\lagrange_1[p,\nu,c,k]$ keeping $k=0$; this relation will come in handy later, because it allows us to work with $\lagrange_1$ and at the very last step set $k=0$ to obtain $\Ec_0$.

We are now left to find a local minimum of $\lagrange_1$ and the easiest way to do this is by looking at its functional derivative w.r.t. $p$. Indeed, recall that the functional derivative of $\lagrange_{1}$ at $p\in\domain_{r}$ is the distribution $\delta\lagrange_{1}[p, \nu, c, k]$ whose action on test functions $\phi\in\tame$ is given by:\footnote{Since $\tame$ is a locally convex space, this is just another guise of the G\^ateaux/Fr\'echet derivative.}
\begin{equation}
\label{eq:funcderiv}
\left\langle\delta\lagrange_{1}[p],\phi\right\rangle
			= \frac{d}{dt}\bigg|_{t=0} \!\!\lagrange_{1}[p + t\phi].
\end{equation}
Note now that the expression $\lagrange_{1}[p+t\phi]$ is well-defined for all $p\in\domain_{r}$, $\phi\in\tame$, thanks to lemma \ref{lem:tame} so that, at least, it makes sense to study its behavior as $t\to0$. In addition to that, our convexity result (lemma \ref{lem:convex}) simplifies things even more because, if $\delta\lagrange_{1}[p]=0$ for some $p\in\domain_{r}$, it immediately follows that $\lagrange_{1}$ will be attaining its global minimum at $p$.\footnote{Indeed, note that the function $w(t) = \lagrange_{1}[p + t(q-p)], t\in[0,1]$ is strictly convex in $[0,1]$ for any choice of $p$ and $q$ in $\domain_{r}$. Thus, if there were some $q\in\domain_{r}$ with $\lagrange_{1}[q]<\lagrange_{1}[p]$, we would have $w'(0) = 0$ (on account of (\ref{eq:funcderiv})) but also $w(0)>w(1)$, a contradiction.}
Then, maximizing the result with respect to $k$ and $c$ simply corresponds to enforcing the normalization and mutual information constraints that appear in (\ref{eq:L0}) and (\ref{eq:L1}):
\begin{eqnarray}
\label{eq:norm_condition}
\int_0^\infty p(x)\dd x &=& 1 \\
\int_0^\infty p(x) \log(1+\rho x)\dd x &=& r
\label{eq:mutual_information_condition}
\end{eqnarray}
Furthermore, we must also maximize with respect to $\nu$, in order to ensure that $p(x)$ be non-negative in $\RR_{+}$. This optimization constraint can be enforced by observing that $\nu(x)=0$ when $p(x)>0$ and vice-versa, as we shall see below.

As a result, once we manage to find a solution to the above optimization problem, we will have:
\begin{proposition}[Uniqueness of Solution]
\label{prop:uniquenes}
Assume that the tame probability measure $p$ satisfies the stationarity condition:
\begin{equation}
\label{eq:delta_L_cond}
\delta\lagrange[p]=0 \quad\text{(resp. }\delta\lagrange_{1}[p]=0)
\end{equation}
along with the constraint (\ref{eq:norm_condition}) (resp. (\ref{eq:norm_condition}), (\ref{eq:mutual_information_condition})). Then, $p$ is the unique global minimum point of (\ref{eq:E0=infE}) (resp. (\ref{eq:E1=infE})).
\end{proposition}
This proposition stems directly from the convexity of $\energy$ and will be of considerable help to us in what follows because it ensures that any stationary point of $\lagrange,\lagrange_{1}$ which satisfies the relevant constraints will be the (unique) solution to our original minimization problem.

\subsection{Solving the Integral Equation}
\label{sec:integral_equation}

Our task now will be to actually {\it find} the solution of (\ref{eq:funcderiv}), subject to the constraints (\ref{eq:norm_condition}), (\ref{eq:mutual_information_condition}). The solution  for $\energy_0$ in (\ref{eq:minimum_0}) can then be obtained by relaxing the constraint (\ref{eq:mutual_information_condition}) and setting $k=0$ in the final result.
To that end, a brief calculation (see appendix \ref{app:solution}) for the functional derivative for the functional derivative $\delta\lagrange_{1}[p]$ of (\ref{eq:funcderiv}) yields the integral equation:
\begin{eqnarray}
\label{eq:func_deriv_E1_result}
  2\int_0^\infty p(x')\log|x-x'| \dd x'&=&  x - (\beta-1)\log x \\  \nonumber
  &-& c - k\log(1+\rho x)-\nu(x).
\end{eqnarray}
The role of $\nu(x)$ in the above equation is to enforce the inequality constraint $p(x)\geq 0$ for all $x\geq 0$. It is well known \cite{Boyd_book} that $\nu(x)>0$ only when the probability density $p(x)$ vanishes, while when the probability density is positive, $\nu(x)$ has to be zero.

The solution of the integral equation involves the inversion of the integral operator in the left-hand-side of (\ref{eq:func_deriv_E1_result}), which is no simple task, because the inversion process depends on the support $\supp(p)$ of the density $p(x)$\cite{Tricomi_book_IntegralEquations}. As discussed in the previous subsection (and with a fair amount of hindsight gained from the Coulomb gas analogy), we will be looking for compactly supported solutions that are continuous in $(0,\infty)$; in other words, we will be assuming that $\supp(p) = [a,b]$ where $0\leq a <b<\infty$.

There is one important issue that must be mentioned here: when the dimensions of the channel matrix attain the critical value $\beta=1$, we will see that $p$ exhibits two different behaviors depending on the values of $r$ and $\rho$ in constraint (\ref{eq:mutual_information_condition}). On one hand, we could have $a>0$ which, by continuity, introduces the constraint $p(a)=0$; on the other hand, we could also have solutions with $a=0$ (which impose no extra constraints because $p$ is assumed continuous only on $(0,\infty)$). If the rate $r$ is less than some critical value $r_{c}(\rho)$, it turns out that solutions with $a>0$ must be rejected because they attain negative values. In that case, we are led to solutions with $a=0$ which have no such problems; the converse happens when $r>r_{c}$, while when $r=r_{c}$ the two solutions coincide.

%


Having said that, we may return to (\ref{eq:func_deriv_E1_result}), where we have $\nu(x)>0$ if and only if $p(x)=0$. By restricting $x$ to lie in the interval $[a,b]$, we may henceforth ignore $\nu(x)$ altogether. Furthermore, to eliminate $c$ for the moment, a differentiation of (\ref{eq:func_deriv_E1_result}) with respect to $x$ yields:
\begin{equation}
\label{eq:func_deriv_E1_result_3}
2{\cal P} \int_a^b \frac{p(x')}{x-x'} dx' = 1-\frac{\beta-1}{x}-\frac{k\rho}{1+\rho x} \equiv f(x)
\end{equation}
where ${\cal P}$ denotes the Cauchy principal value of the integral.\footnote{The principle value appears because of the absolute value $|x'-x|$ in (\ref{eq:func_deriv_E1_result}).}

The above equation has a straightforward physical meaning: it represents a balance of forces at every location $a\leq x <b$, because the repulsion from all other charges of the distribution located at $x'$ (the LHS expression) is equal to the external forces (RHS). For $\beta>1$, we intuitively expect that $p(x)$ must vanish at $x=0$ because in this case the force from the finite charge density located at $x=0$  (the second term of (\ref{eq:func_deriv_E1_result_3})) would be infinite. As a result, we intuitively expect that $a>0$ for all $\beta>1$; this expectation will be vindicated shortly.

Indeed, the solution of this integral equation for general $f(x)$ can be obtained using standard methods from the theory of integral equations \cite{Mikhlin_book_IntegralEquations, Tricomi_book_IntegralEquations}. So as not to interrupt the presentation, we will postpone the details until appendix \ref{app:solution} and will only give the final result here:
\begin{eqnarray}
\label{eq:gen_solution_int_eq0}
    p(x) &=& \frac{{\cal P} \int_a^b  \frac{\sqrt{(y-a)(b-y)}f(y)}{y-x} dy + C'}{2\pi^2\sqrt{(x-a)(b-x)}} \\ \nonumber
    &=& \frac{-x - \frac{k\sqrt{(1+a\rho)(1+b\rho)}}{1+\rho x} -\frac{(\beta-1)\sqrt{ab}}{x} + C}{2\pi\sqrt{(x-a)(b-x)}}
\end{eqnarray}
where $C,C'$ are unknown constants to be determined by the condition $p(b)=0$.

As we explain in Appendix \ref{app:solution}, this formula is valid only when the function $f$ is itself $L^{\eta}$-integrable for some $\eta>1$. This is always true if $\beta=1$, because the singular term proportional to $(\beta-1)$ is not present in the LHS of (\ref{eq:func_deriv_E1_result_3}). However, as we have already mentioned, the case $\beta=1$ has its own set of subtleties, analyzed at length in section \ref{sec:beta=1}. In particular, we obtain two different solutions depending on whether the support of $p$ extends to $0$ or not (imposing the constraints $a=0$ or $p(a)=0$ respectively), but only one of them is physically admissible (i.e. is a tame probability measure lying in the rate-constrained domain $\domain_{r}$).

On the other hand, this dichotomy ceases to exist when $\beta>1$. Indeed, if $\beta>1$ and $a=0$, the LHS of (\ref{eq:func_deriv_E1_result_3}) is no longer integrable. However, the RHS of (\ref{eq:func_deriv_E1_result_3}) {\it is} $L^{1+\eps}$-integrable whenever $p$ is itself $\eps$-tame, on account of the properties of the finite Hilbert transform \cite{Tricomi_book_IntegralEquations} (see also appendix \ref{app:solution}). We thus conclude that any solution to (\ref{eq:func_deriv_E1_result_3}) whose support extends to $0$ cannot be tame and will thus have to be rejected. As a result, the support of $p$ for $\beta>1$ has to be bounded away from $0$, thus leading to the constraint $p(a)=0$ and proving our intuitive expectation above.

%


So, starting with the general case $a,b>0$, we find that the constraint of continuity requires that the distribution $p(x)$ vanish at the endpoints $a,b$ of its support. The condition $p(b)=0$ determines the value of $C$ in (\ref{eq:gen_solution_int_eq0}) resulting in the following form for $p(x)$:
\begin{equation}
\label{eq:gen_solution_int_eq1}
p(x)	= \frac{\sqrt{b-x}}{2\pi\sqrt{x-a}} \left(1 - \frac{k\rho}{(1+\rho x)}\sqrt{\frac{1+a\rho}{1+b\rho}} -\frac{\beta-1}{x}\sqrt{\frac{a}{b}}\right)
\end{equation}
The additional condition $p(a)=0$ (when $a>0$) results to
\begin{eqnarray}\label{eq:p_x_beta>1}
    p(x) &=& \frac{1}{2\pi} \frac{\sqrt{(b-x)(x-a)}}{x(1+\rho x)} \left(\rho x+\frac{\beta-1}{\sqrt{ab}}\right)
\end{eqnarray}
with the value of $a$ determined (as a  function of $b$ and $k$) by the equation:
\begin{equation}\label{eq:p_a_=0}
    \frac{k\rho}{\sqrt{(1+\rho a)(1+ \rho b)}} + \frac{\beta-1}{\sqrt{ab}} = 1.
\end{equation}
Demanding that $p$ be properly normalized as in (\ref{eq:norm_condition}), imposes the constraint:
\begin{eqnarray}\label{eq:norm_integral}
\int_a^b p(x) dx &=& \frac{a+b-2k-2(\beta-1)}{4} \\ \nonumber
&+&\frac{k}{2\sqrt{(1+a\rho)(1+b\rho)}} =1.
\end{eqnarray}
In Appendix \ref{app:uniqueness_ab} we show that (\ref{eq:p_a_=0}) and (\ref{eq:norm_integral}) admit a unique solution $a,b$ for any given $k$ and, as a result, Proposition \ref{prop:uniquenes} guarantees the existence of a (necessarily unique) density $p(x)$ that minimizes (\ref{eq:minimum_1}).

Now, given the resulting solution $p(x)$ we can readily calculate the minimum energy $\energy[p]$ itself:
\begin{eqnarray}
\energy[p] &=& \int_a^b x p(x) \dd x - (\beta-1) \int_a^b p(x) \log x \dd x
\nonumber  \\
&-& \int_a^b \int_a^b p(x) p(y) \log|x-y| \dd y \dd x
\nonumber \\
&=& \frac{1}{2}\int_a^b x p(x) dx - \frac{\beta-1}{2} \int_a^b p(x) \log x \dd x
\nonumber \\
&+& \frac{k}{2}\int_a^b p(x) \log(1+\rho x) \dd x +\frac{c}{2}
\label{eq:energy_functional_calc}
\end{eqnarray}
where in the second line we eliminated the double integral by substituting it from (\ref{eq:func_deriv_E1_result}) \cite{Vivo2007_LargeDeviationsWishart}. As for the value of $c$ itself, it can be determined by evaluating (\ref{eq:func_deriv_E1_result}) at a fixed value of $x$, say $x=a$:
\begin{eqnarray}
\label{eq:c_calc}
 c &=& a - (\beta-1)\log a - k \log(1+\rho a) \\ \nonumber
 &-& 2 \int_a^b \log(x-a) p(x) dx
\end{eqnarray}
Inserting this in (\ref{eq:energy_functional_calc}) then yields:
\begin{eqnarray}
\label{eq:energy_functional_calc1}
\energy[p] &=& \frac{1}{2}\int_a^b x p(x) dx - \frac{\beta-1}{2} \int_a^b p(x) \log x dx \\ \nonumber
&-& \int_a^b p(x)\log(x-a) dx \\ \nonumber
&+& \frac{1}{2}\left(k\left(r-\log(1+\rho a)\right)+a-(\beta-1)\log a\right)
\end{eqnarray}

\section{Probability Distributions $P_N(r)$, $P_{\text{out}}(r)$}
\label{sec:results}

The central aim of the paper is to evaluate the probability density of the rate $r$ for large $N$, namely $P_N(r)$ given by (\ref{eq:P_N(r)})
\begin{equation}\label{eq:P_N(r)_norm_def}
    P_N(r) \approx B_N e^{-N^2(\energy_1(r)-\energy_0)}
\end{equation}
where $B_N$ is a normalization constant, while $\energy_1(r)$ (\ref{eq:E1=infE}) and $\energy_0$ (\ref{eq:E0=infE}) are the most probable values of the energy evaluated with and without the mutual information constraint (\ref{eq:mutual_information_condition}), respectively. In this section we will calculate these values and derive the corresponding eigenvalue probability densities $p(x)$ that minimize the energy functional $\energy[p]$. In Section \ref{sec:MP_Distribution}, we will derive $\energy_0$ and we will show how the corresponding density $p(x)$ is the Mar\v{c}enko-Pastur Distribution. In Sections \ref{sec:beta>1} and \ref{sec:beta=1} we will calculate $\energy_1(r)$ for the cases $\beta>1$ and $\beta=1$ respectively. Finally, in Section \ref{sec:OutageProbability} we will show how one can calculate the outage probability $P_{out}(r)$.

\subsection{Evaluation of $\energy_0$}
\label{sec:MP_Distribution}

As mentioned above, it is instructive to first calculate the most probable distribution of eigenvalues without the mutual information constraint (\ref{eq:mutual_information_condition}), which will end up being the well-known Mar\v{c}enko-Pastur distribution. This can be immediately extracted from the analysis in Section \ref{sec:integral_equation} by setting $k=0$. Solving for $a,b$ in (\ref{eq:p_a_=0}), (\ref{eq:norm_integral}) gives
\begin{eqnarray}
\label{eq:a_b_MP}
a &=& \left(\sqrt{\beta}-1\right)^2 \\ \nonumber
b &=& \left(\sqrt{\beta}+1\right)^2
\end{eqnarray}
and (\ref{eq:p_x_beta>1}) then takes the well-known form (\ref{eq:MP_def}). \footnote{Note that when $\beta=1$, the lower endpoint vanishes ($a=0$) and a square-root (integrable) singularity appears in $p(x)$ in (\ref{eq:MP_def}).}

We may also evaluate the energy $\energy_0$ by setting $k=0$ in (\ref{eq:energy_functional_calc1}). Thus we get:
\begin{eqnarray}
\label{eq:energy_functional_MP}
\energy_0 &=& \frac{1}{2}\int_a^b x p(x) \dd x +\frac{1}{2}\left(a-(\beta-1)\log a\right)
\\ \nonumber
&-&\frac{\beta-1}{2} \int_a^b p(x) \log x \dd x - \int_a^b p(x) \log(x-a) \dd x
\end{eqnarray}
and, after some algebra, we can rewrite the above expression in the closed form:
\begin{eqnarray}
\label{eq:energy_functional_MP_closed_form}
\energy_0 &=& \frac{\Delta^2}{32} + \frac{a}{2} -\log\Delta -\frac{\beta-1}{2}\log(a\Delta) \\ \nonumber
&-& \frac{\Delta}{2}\left[G\left(0,\frac{a}{\Delta}\right) + \frac{\beta-1}{2} G\left(\frac{a}{\Delta},\frac{a}{\Delta}\right)\right]
\end{eqnarray}
where $\Delta\equiv b-a$ and the function $G(x,y)$ is given by \cite{Chen1996_EigDistributionsLaguerre}:
\begin{eqnarray}\label{eq:Q_fun}
G(x,y)	&=& \frac{1}{\pi}\int_0^1 \sqrt{t(1-t)} \frac{\log(t+x)}{t+y}\dd t \\ \nonumber
		&=& -2\sqrt{y(1+y)}\log\left[\frac{\sqrt{x(1+y)}+\sqrt{y(1+x)}}{\sqrt{1+y}+\sqrt{y}}\right] \\ \nonumber
  &+& \left(1+2y\right)\log\left[\frac{\sqrt{1+x}+\sqrt{x}}{2}\right] \\ \nonumber
  &-& \frac{1}{2}\left(\sqrt{1+x}-\sqrt{x}\right)^2
\end{eqnarray}
When $\beta=1$, $a$, $b$ in (\ref{eq:a_b_MP}) take the values $b=4$ and $a=0$, and hence (\ref{eq:energy_functional_MP}) becomes $\energy_0=3/2$.

\subsection{Evaluation of $\energy_1(r)$: $\beta>1$}
\label{sec:beta>1}

We will now calculate $\energy_1$ for the case $\beta>1$. To do so, we need to evaluate the constants $a$, $b$, $k$ as a function of $r$ and $\rho$ using (\ref{eq:p_a_=0}), (\ref{eq:norm_integral}) and (\ref{eq:mutual_information_condition}). The values of these constants will determine the density of eigenvalues constrained on the subset with fixed total rate $R=Nr$ in the large $N$ limit.  After inserting (\ref{eq:p_x_beta>1}) into the last equation and integrating, (\ref{eq:mutual_information_condition}) can be expressed explicitly as
\begin{eqnarray}
\label{eq:mut_info_beta>1}
r& =& \int_a^b p(x) \log(1+\rho x) \dd x\\ \nonumber
&=& \log\Delta\rho
+ \frac{\Delta k\rho}{2\sqrt{(1+\rho a)(1+\rho b)}} G\left(\frac{1+\rho a}{\Delta\rho},\frac{1+\rho a}{\Delta\rho}\right) \\ \nonumber
&+& \frac{\Delta}{2}\left(1-\frac{k\rho}{\sqrt{(1+\rho a)(1+\rho b)}}\right) G\left(\frac{1+\rho a}{\Delta\rho},\frac{a}{\Delta}\right)
\end{eqnarray}
where $G(x,y)$ is given in (\ref{eq:Q_fun}).

Based on the arguments discussed in the previous section, it suffices to show that there exists a distribution $p(x)$ in the form of (\ref{eq:p_x_beta>1}) satisfying the constraints (\ref{eq:norm_condition}), (\ref{eq:mutual_information_condition}). This corresponds to finding values of $a$, $b$, and $k$ that satisfy (\ref{eq:p_a_=0}), (\ref{eq:norm_integral}) and (\ref{eq:mut_info_beta>1}), while at the same time maintaining $p(x)\geq 0$ for all $x\in [a,b]$. If such a solution exists, then according to Theorem \ref{prop:uniquenes} it will be unique.

In Appendix \ref{app:uniqueness_ab} we show that equations (\ref{eq:p_a_=0}) and (\ref{eq:norm_integral}) admit a unique solution for any $k$. We therefore only need to show that (\ref{eq:mut_info_beta>1}) has a solution in $k$ for any $r>0$.

It suffices to show that the function defined solely as a function of $k$ by the right-hand-side of (\ref{eq:mut_info_beta>1}) (with $a$ and $b$ expressed in terms of $k$) takes all values in $(0,\infty)$. Hence by continuity it will attain the value $r$ for all positive rates $r>0$.
We first see that as $k\rightarrow -\infty$ the solution of (\ref{eq:p_a_=0}), (\ref{eq:norm_integral}) is $a\approx (\sqrt{\beta}-1)^2/(\rho|k|)$ and $b\approx (\sqrt{\beta}+1)^2/(\rho |k|)$; then, inserting these solutions into (\ref{eq:mut_info_beta>1}), we see that it may be written in leading order as $r\approx \beta/|k|$. On the other hand, for $k\rightarrow \infty$ (\ref{eq:p_a_=0}), (\ref{eq:norm_integral}) give $a\approx \sqrt{k+\beta-\rho^{-1}/2}-1$ and $b\approx \sqrt{k+\beta-\rho^{-1}/2}+1$, resulting to $r\approx \log k\rho$. This shows that the corresponding solution $p(x;r)$ is the unique minimizing distribution of $\energy$ in $\domain_r$.

\begin{figure}[htb]
\centerline{\epsfxsize=1.0\columnwidth\epsffile{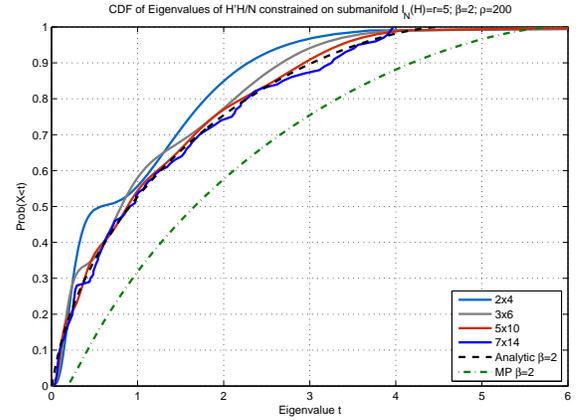}}
\caption{Cumulative distribution function (CDF) of eigenvalues for a conventional and a generalized {\sf MP} distribution with $\beta=2$, $r=5$ and $\rho=200$. For $\beta=2$ and $\rho=100$, the value of the ergodic mutual information is $r_{erg}=5.0014$. Thus, the generalized {\sf MP} distribution with $\rho=100$ would correspond to the conventional {\sf MP} distribution above. Also plotted are the empirical CDFs for eigenvalues of $\bH^\dagger\bH/N$ conditioned on the subset $I_N\leq Nr$. It is remarkable that even for the not-too-large antenna array system $5\times 10$ the empirical distribution converges to the analytic result.} \label{fig:gen_mp_beta_2_comp}
\end{figure}

\begin{figure*}
\centerline{%
\subfigure[Fixed $r=5.78$]
{\epsfxsize=.49\textwidth
 \epsffile{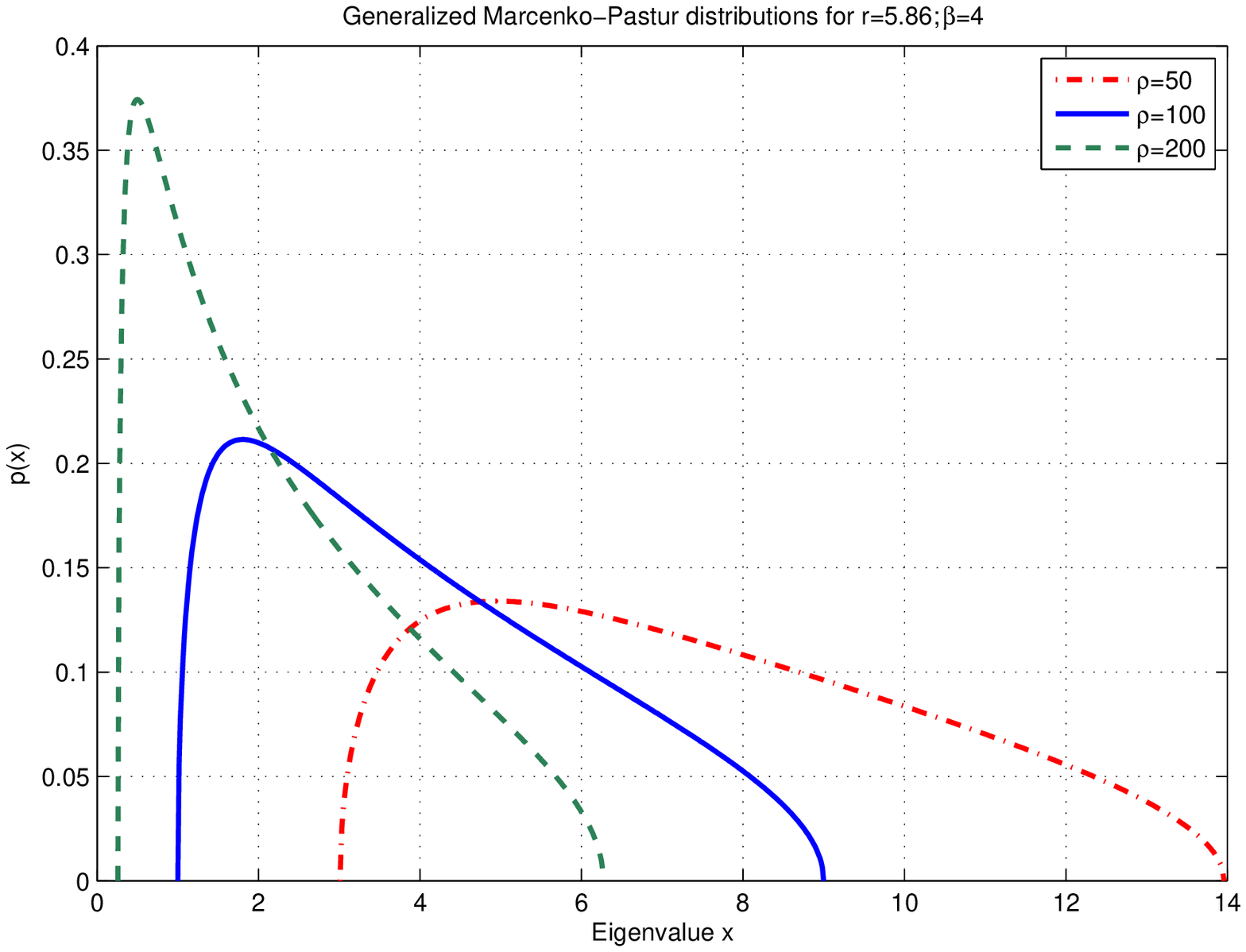} }
\hfil %
\subfigure[Fixed $\rho=100$ ]
{\epsfxsize=.49\textwidth
 \epsffile{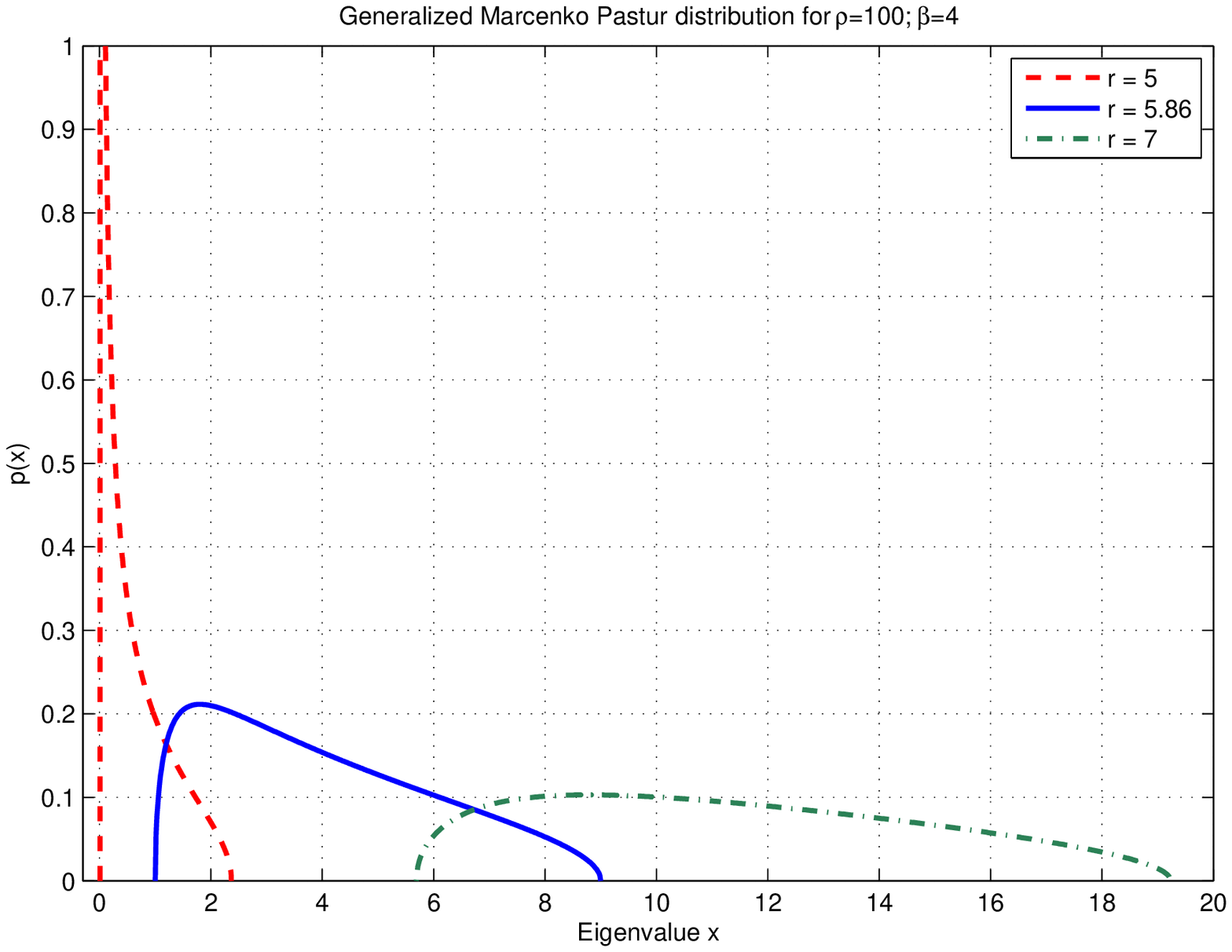}}}
\caption{Generalized {\sf MP} distributions for $\beta=4$ and different values of $\rho$ and $r$. In (a) we plot the eigenvalue distributions for different values of $\rho$ and fixed $r=5.78$, which is the value of $r_{erg}$ for the curve in the middle with $\rho=100$. In (b) we plot the eigenvalue distributions for fixed $\rho=100$ and different values $r$. We see that in the latter plot the distribution is more sensitive on $r$ rather than $\rho$.} \label{fig:gen_mp_beta_4}
\end{figure*}

In Fig. \ref{fig:gen_mp_beta_2_comp} we compare this distribution with the corresponding empirical probability distribution function obtained by numerical simulations. We see that the agreement is quite remarkable, indicating a quick convergence to the asymptotic distribution function of the eigenvalues constrained at the tails of the distribution of the mutual information. Furthermore, to get a feeling for the dependence of the eigenvalue distributions in terms of their parameters, in Fig. \ref{fig:gen_mp_beta_4} we plot a few representative examples.

We may now calculate the value of $\energy_1$. Inserting $p(x)$  from (\ref{eq:p_x_beta>1}) into (\ref{eq:energy_functional_calc1}) and integrating finally gives us:
\begin{eqnarray}
\label{eq:E1_beta>1}
\energy_1 &=& \frac{\Delta^2}{32} + \frac{a}{2} -\log\Delta -\frac{\beta-1}{2}\log(a\Delta)
\\ \nonumber
&+&\frac{k}{2}\left(r-\log(1+\rho a)-\frac{\left(\sqrt{1+\rho b}-\sqrt{1+\rho a}\right)^2}{4\rho\sqrt{(1+\rho a)(1+\rho b)}}\right)
\\ \nonumber
&-& \frac{\Delta k\rho}{2\sqrt{(1+\rho a)(1+\rho b)}} \\ \nonumber
&\cdot& \left[ G\left(0,\frac{1+\rho a}{\Delta\rho}\right)+ \frac{\beta-1}{2}G\left(\frac{a}{\Delta},\frac{1+\rho a}{\Delta\rho}\right)\right] \\ \nonumber
&-& \frac{\Delta}{2}\left(1-\frac{k\rho}{\sqrt{(1+\rho a)(1+\rho b)}}\right) \\ \nonumber
&\cdot& \left[ G\left(0,\frac{a}{\Delta}\right) +\frac{\beta-1}{2}  G\left(\frac{a}{\Delta},\frac{a}{\Delta}\right) \right]
\end{eqnarray}
where $G(x,y)$ is given by (\ref{eq:Q_fun}).
Plugging this together with $\energy_0$ into (\ref{eq:P_N(r)_norm_def}) we obtain $P_N(r)$, up to the normalization constant.

\subsection{Evaluation of $\energy_1(r)$: $\beta=1$}
\label{sec:beta=1}

The case $\beta=1$ deserves special attention. In this case the logarithmic repulsion from the $\delta$-function density of eigenvalues at the origin in (\ref{eq:energy}) and (\ref{eq:func_deriv_E1_result}) is no longer present. As discussed in Section \ref{sec:methodology}.B, depending on the parameters $r$ and $\rho$ there are two distinct types of solutions, which we treat here separately.

\subsubsection{Case $\beta=1$ and $r>r_c(\rho)$}
\label{sec:case r>r_c}

We start by attempting to solve the problem as in the $\beta>1$ case, namely by looking for solutions of $0<a<b$ for the distribution's support. It is straightforward to show that
the conditions (\ref{eq:p_a_=0}) and (\ref{eq:norm_condition}) yield the following values for $a$, $b$ when $\beta=1$:
\begin{eqnarray}\label{eq:a_b_beta=1_r>r_c}
a &=&  \left(\sqrt{k+1}-1\right)^2 - \rho^{-1} \nonumber \\
b &=&   \left(\sqrt{k+1}+1\right)^2 - \rho^{-1}.
\end{eqnarray}
As a result, the probality density function $p$ becomes:
\begin{equation}\label{eq:p_x_beta=1_r>rc}
    p(x) = \frac{\rho}{2\pi}\frac{\sqrt{(b-x)(x-a)}}{1+\rho x}
\end{equation}

The value of the parameter $k$ can be obtained in a unique way from the mutual information condition, which now reads:
\begin{equation}\label{eq:mut_info_beta=1_r>rc}
r-\log\rho=(k+1)\log(k+1)-k\log k-1.
\end{equation}
The monotonicity of the right-hand-side of this equation with respect to $k$ implies a unique $k(r)$ satisfying (\ref{eq:mut_info_beta=1_r>rc}) and hence a unique set of $a$,$b$ in (\ref{eq:a_b_beta=1_r>r_c}), guaranteeing uniqueness of (\ref{eq:p_x_beta=1_r>rc}).

In its turn, this can be used to evaluate the value of the outage exponent:
\begin{equation}\label{eq:E1_beta=1_r>rc}
\Ec_1-\Ec_0= \frac{k-1}{2}(r-\log\rho) +k-\frac{1}{2}-\rho^{-1}-\frac{k\log k}{2}.
\end{equation}

From (\ref{eq:a_b_beta=1_r>r_c}) we can see that this solution can only be valid for $k\geq k_c(z)\equiv \rho^{-1}+2/\sqrt{\rho}$, or equivalently for $r>r_c(\rho)$ where
\begin{eqnarray}
\label{eq:r_c}
    r_c(\rho) &\equiv& \frac{1+2\sqrt{\rho}}{\rho}\log\left(1+\frac{\rho}{1+2\rho}\right) \\ \nonumber
    &+&2\log\left(1+\sqrt{\rho}\right)-1 >r_{\text{erg}}
\end{eqnarray}
The reason is that for  $k<k_c(\rho)$ (or $r<r_c(\rho)$) the value of $a$ becomes negative, which is unacceptable.

\subsubsection{Case $\beta=1$ and $r\leq r_c(\rho)$}
\label{sec:case r<r_c}

In this case we can no longer treat $a$ as a free variable. Instead, because $p(x)=0$ for $x<0$, the charge density becomes confined at the boundary $x=0$. Thus, we need to look for solutions of (\ref{eq:func_deriv_E1_result}) with $a=0$, in which case the charge density has a square-root singularity at $x=0$ (instead of vanishing continuously). This is actually quite natural since we expect that, for $k=0$ (or, equivalently, for $r=r_{\text{erg}}$), the charge distribution should take the form of the $\beta=1$ Mar\v{c}enko-Pastur density:
\begin{equation}
\label{eq:p_x_beta=1_MP}
    p(x) = \frac{\sqrt{4-x}}{2\pi\sqrt{x}}.
\end{equation}

Indeed, for general $b$, $k$, the distribution becomes:
\begin{equation}
\label{eq:p_x_beta=1_r<r_c}
    p(x) = \frac{\sqrt{b-x}}{2\pi(1+\rho x)\sqrt{x}}\left(\rho x+1-\frac{k\rho }{\sqrt{1+\rho b}}\right),
\end{equation}
and the normalization condition (\ref{eq:norm_condition}) implies
\begin{equation}\label{eq:norm_beta=1_r<r_c}
   k= \frac{\frac{b}{2}-2}{1-\frac{1}{\sqrt{1+\rho b}}}
\end{equation}
It can easily be shown that the right-hand-side of (\ref{eq:norm_beta=1_r<r_c}) is increasing in $b$ and, hence, (\ref{eq:norm_beta=1_r<r_c}) has a unique solution in $b$ for all $k$.

In the last case ($a=0$), the mutual information condition (\ref{eq:mutual_information_condition}) can be integrated using (\ref{eq:p_x_beta=1_r<r_c}) to give:
\begin{eqnarray}
\label{eq:mut_info_beta=1_r<r_c}
r&=& 2(k+1)\log\frac{1+\sqrt{1+\rho b}}{2} \nonumber \\
&-&\frac{1}{4\rho}\left(\sqrt{1+\rho b}-1\right)^2-\frac{k}{2}\log\left(1+\rho b\right).
\end{eqnarray}
We may use the same argument as in the previous subsection to show that this equation has at least one solution for any $0<r<r_c(\rho)$. Indeed when $k=k_c$, the right-hand-side above takes the value of $r_c$. In contrast, when $k\rightarrow -\infty$, (\ref{eq:norm_beta=1_r<r_c}) gives $b\approx 4/(\rho |k|)$, in which case the right-hand-side of (\ref{eq:mut_info_beta=1_r<r_c}) becomes $\approx 1/|k|$. Thus all values between $(0, r_c(\rho))$ are taken when $k\in (-\infty,k_c(\rho))$. Hence by continuity it will attain the value $r\in (0,r_c)$.

After solving for $b$ and $k$ as a function of $r$ and $\rho$, $\energy_1$ can be calculated easily. Therefore, the exponent of the probability distribution $P_N(r)$ becomes:
\begin{eqnarray}\label{eq:E1_beta=1_r<rc}
\Ec_1-\Ec_0 &=& \frac{k}{2}\left(r-\frac{b}{4}\right)- \log\frac{b}{4} - k \log\frac{1+\sqrt{1+\rho b}}{2}  \nonumber \\
    &+& \frac{1}{32}\left(b - 4\right)\left(4\rho^{-1}+3b+12\right)
\end{eqnarray}

We should point out that just as the solution (\ref{eq:p_x_beta=1_r<r_c}) is not valid for $r>r_c(\rho)$, the solution (\ref{eq:p_x_beta=1_r<r_c}), which we found to be valid for $r>r_c(\rho)$ is not valid for $r<r_c(\rho)$. To see this, it is straightforward to show that in this case the constant term in the last parenthesis in (\ref{eq:p_x_beta=1_r<r_c}) (namely $1-k\rho/\sqrt{1+\rho b}$) is negative. As a result, (\ref{eq:p_x_beta=1_r<r_c}) cannot be valid for $k<k_c(\rho)$ because the charge density becomes negative at some point $x>0$. As a result the solutions we found above are unique in their domains of validity. Interestingly there is a weak, third order discontinuity at the transition $r=r_c(\rho)$, in the sense that the first two derivatives of $\energy_1(r)$ with respect to $r$ evaluated at $r=r_c$ are continuous, while the third is discontinuous. This is analogous to the phase transition observed in \cite{Vivo2008_DistributionsConductanceShotNoise}.


\subsection{Evaluation of the Outage Probability $P_{out}(r)$}
\label{sec:OutageProbability}

In this section we will calculate the outage probability $P_{out}(r)=\prob(I_N<Nr)$ from  $\energy_1(r)$. To do this we need to integrate $\exp\left[-N^2(\Ec_1(r)-\Ec_0)\right]$ over $r$. Generally it is impossible to evaluate  this integral in closed form. Nevertheless, due to the presence of the factor $N$ in the exponent, $P_N(r)$ falls rapidly away from its peak and thus we may use Watson's lemma\cite{Bender_Orszag_book} (a special case of Varadhan's lemma), to evaluate the asymptotic value of the integral. First, we will calculate the normalization factor of the distribution.

As we shall see in Section \ref{sec:limiting_cases} for $r$ close to $r_{\text{erg}}$, $\Ec_1(r)-\Ec_0\sim (r-r_{\text{erg}})^2/v_{\text{erg}}$, where $v_{\text{erg}}$ is the ergodic variance (\ref{eq:var_erg_all_alpha}) of the mutual information distribution. Therefore, we have
\begin{equation}\label{eq:Watson_lemma_int_P(R)}
\int_0^\infty e^{-N^2(\Ec_1(r)-\Ec_0)} \dd r \approx \int_0^\infty e^{-\frac{N^2(r-r_{\text{erg}})^2}{2v_{\text{erg}}}} \dd r \approx \frac{\sqrt{2\pi v_{\text{erg}}}}{N}
\end{equation}
which then gives
\begin{equation}\label{eq:P_R_result}
    P_N(r) \approx \frac{N}{\sqrt{2\pi v_{\text{erg}}}} e^{-N^2\left(\Ec_1(r)-\Ec_0\right)}
\end{equation}
and fixes the normalization constant in (\ref{eq:P_N(r)_norm_def}).
To calculate the outage probability $\outage(r)=\prob(I_N<Nr)$ to leading order in $N$, we first note that for $r<r_{\text{erg}}$ ($r>r_{\text{erg}}$), $\Ec_1(r)$ is a decreasing (increasing) function of $r$. Therefore, to leading order, the behavior will be dominated by the value of the exponent at $r$. Using Watson's lemma once again we obtain the following expression for the outage probability:
\begin{equation}\label{eq:outage_prob}
    P_{out}(r) \approx
    \frac{e^{-N^2\left[\Ec_1(r)-\Ec_0-\frac{\Ec_1'(r)^2}{2\Ec_1''(r)}\right]}Q\left(\frac{N\left|\Ec_1'(r)\right|}{\sqrt{\Ec_1''(r)}} \right)      }{\sqrt{\Ec_1''(r) v_{erg}}}
\end{equation}
when $r<r_{\text{erg}}$ and
\begin{equation}\label{eq:outage_prob2}
    P_{out}(r) \approx
    1 - \frac{e^{-N^2\left[\Ec_1(r)-\Ec_0-\frac{\Ec_1'(r)^2}{2\Ec_1''(r)}\right]}Q\left[\frac{N\left|\Ec_1'(r)\right|}{\sqrt{\Ec_1''(r)}} \right] }{\sqrt{\Ec_1''(r) v_{erg}}}
\end{equation}
when $r>r_{\text{erg}}$.
In the above, $\Ec_1'(r)$ and $\Ec_1''(r)$ are the first and second derivatives of $\Ec_1(r)$ with respect to $r$ and $Q(x)$ is given by
\begin{equation}\label{eq:Qx_def}
    Q(x) = \int_x^\infty \frac{dx}{\sqrt{2\pi}} e^{-\frac{t^2}{2}}
\end{equation}


\begin{figure}[htb]
\centerline{\epsfxsize=1.0\columnwidth\epsffile{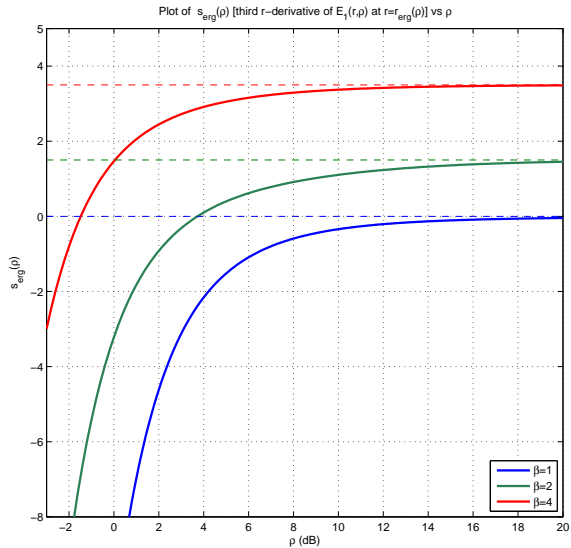}}
\caption{Dependence of  $s_{erg}=\energy_1'''(r_{erg})$ on $\rho$ for different values of $\beta$. We see that for not too large $\rho$ the behavior of $s_{erg}$ quickly converges to the correct asymptotic limit (\ref{eq:E3_asymptotics}), included here with dashed lines. } \label{fig:E3_vs_rho}
\end{figure}


\section{Analysis of Limiting Cases}
\label{sec:limiting_cases}

We will now analyze the results of the previous section in specific limiting cases of the parameter space $(\rho, r, \beta)$. We will thereby be able to connect  with already existing results in specific regions, and also to describe the behavior of the probability density of $P_N(r)$ in other regions, which hitherto have defied asymptotic analysis.

\subsection{Gaussian Region $r\approx r_{\text{erg}}(\rho)$}
\label{sec:GaussianApprox}

The most relevant limiting case is the Gaussian regime: after all, the Gaussian approximation, as well as the present approach assume that the number of antennas $N$ is large. The difference is that our approach does not focus only in the region of $N|r-r_{\text{erg}}|=\bigoh(1)$, where the Gaussian approximation should be valid. To reach that limit, we need to analyze the small $k$ region of (\ref{eq:mut_info_beta>1}), (\ref{eq:mut_info_beta=1_r<r_c}) since,  in the limit $k=0$, both equations reduce to $r=r_{\text{erg}}(\rho)$, where the normalized ergodic mutual information $r_{\text{erg}}$ is well known to be \cite{Moustakas2003_MIMO1, Verdu1999_MIMO1, Rapajic2000_InfoCapacityOfARandomSignatureMIMOChannel}:
\begin{eqnarray}
\label{eq:r_erg_all_alpha}
  r_{\text{erg}} =  \log u + \beta \log\left[1+\frac{\rho}{u}\right] - \left(1-u^{-1}\right)
\end{eqnarray}
with:
\begin{equation}\label{eq:u_def}
    u = \frac{1}{2}\left(1+\rho(\beta-1)+\sqrt{(1+\rho(\beta-1))^2+4\rho}\right)
\end{equation}
By implicitly differentiating $a$, $b$, $k$ with respect to $r$ through the equations that define them, and expressing their values and the values of their derivatives at $r=r_{erg}$ we can obtain the following expansion
\begin{equation}\label{eq:E1-E0_small_k2}
    \energy_1-\energy_0 = \frac{\left(r-r_{\text{erg}}\right)^2}{2v_{\text{erg}}} + \frac{s_{erg}}{6}\left(r-r_{\text{erg}}\right)^3 + \bigoh\left(\left(r-r_{\text{erg}}\right)^4\right)
\end{equation}
where
\begin{eqnarray}
\label{eq:var_erg_all_alpha}
  v_{\text{erg}} =  -\log\left[1-\frac{(1-u)^2}{\beta u^2}\right]
\end{eqnarray}
coincides with the variance of the mutual information distribution as analyzed in \cite{Moustakas2003_MIMO1, Hachem2006_GaussianCapacityKroneckerProduct},
and $s_{erg}$ is the third {\it total} derivative of $\energy_1$ with respect to $r$ and evaluated at $r=r_{erg}(\rho)$. Without the cubic term, (\ref{eq:E1-E0_small_k2}) is exactly the Gaussian limit of the mutual information distribution discussed in various papers in the past. This Gaussian limit is valid as long as the cubic (as well as all higher order) terms in the exponent of the probability are smaller than unity. Since this condition depends on $s_{erg}$, it is worth looking its behavior with $\rho$. In Fig. \ref{fig:E3_vs_rho} we plot $s_{erg}$ as a function of $\rho$. We see that it has a well-defined limit for large $\rho$. Specifically, it has the following asymptotic form
\begin{equation}\label{eq:E3_asymptotics}
    s_{erg}(\rho)\approx \left\{
                           \begin{array}{cc}
                             -\frac{2}{\log(\rho)^3} & \beta=1 \\
                             -\frac{1}{\beta(\beta-1)\log\left(1-\beta^{-1}\right)^3} & \beta>1 \\
                           \end{array}
                         \right.
\end{equation}
Also, for small $\rho\ll 1$ we can show that $s_{erg}\approx -c_\beta/\rho^3$, where $c_{\beta}>0$ is a constant that depends on $\beta$. Thus the condition for validity of the Gaussian approximation is
\begin{equation}\label{eq:Gaussian_validity}
    \left|r-r_{erg}(\rho)\right| \ll \sqrt[3]{\frac{6}{\left|s_{erg}\right|}} N^{-2/3}
\end{equation}
We therefore see that the Gaussian approximation should not be valid for significant deviations from $r_{erg}$, e.g. $r=r_{erg}/2$. In contrast our large deviations (LD) approximation continues to be valid in that rate region as well.

\subsection{Large $\rho$ Approximation: $r<r_{\text{erg}}$}
\label{sec:DMT_region_r<rerg}

Next we analyze the behavior of the probability distribution of $r$ in the large $\rho$ limit, while keeping the ratio $r/\log\rho$ finite and less than $1$.\footnote{This is the region analyzed in the diversity-multiplexing trade-off \cite{Zheng2003_DiversityMultiplexing}.}  Since in the large $\rho$ limit $r_{\text{erg}}\sim \log\rho$, the region $q\leq 1$ with $\rho\gg 1$ corresponds to $k<0$, equations (\ref{eq:norm_integral}), (\ref{eq:p_a_=0}) will admit the following solutions for $a,b$:
\begin{eqnarray}
\label{eq:a_dmt_regime}
  a &\sim& \frac{(\beta-1)^2}{4\rho(1-q)(\beta-q)} \\
\label{eq:b_dmt_regime}
  b & \sim& 4q
\end{eqnarray}
where $q=r/\log\rho$ and we are assuming that $0<q<1$.

Now, note that the lower end of the spectrum has become of order $O(1/\rho)$, while the upper limit is still finite, just as expected. It is also interesting to calculate the proportion of eigenvalues that are in the neighborhood of $x=1/\rho$ when $\rho\rightarrow \infty$. Indeed, by integrating the probability distribution $p(x)$ (\ref{eq:p_x_beta>1}) from $a=\bigoh(\rho^{-1})$ (\ref{eq:a_dmt_regime}) to $L\rho^{-1}$ for some (arbitrarily) large $L$ we get
\begin{eqnarray}
\label{eq:probability_x=O(z)}
  \lim_{L\rightarrow \infty} \lim_{\rho\rightarrow \infty} \prob(\rho x<L) = 1-q
\end{eqnarray}
Thus, the proportion of ``small'' eigenvalues is simply $1-q$, in agreement with \cite{Zheng2003_DiversityMultiplexing}. Plugging (\ref{eq:a_dmt_regime}), (\ref{eq:b_dmt_regime}) into the equation for $\energy_1$ then gives the expected result for the exponent:
\begin{equation}\label{eq:E1-E0_dmt}
    \energy_1-\energy_0 \sim \log\rho\left[(1-q)(\beta-q)\right]
\end{equation}
which is exactly the diversity exponent (divided by $N^2$) of \cite{Zheng2003_DiversityMultiplexing}.

From the above, we see the difference between the two asymptotic analyses discussed above. In the previous section, the eigenvalue distribution did not deviate significantly from the most probable Mar\v{c}enko-Pastur distribution, since $k$ was assumed to be small. In contrast, here, $k$ is finite, and in particular equal to $k=2q-1-\beta$, In addition, a significant portion of the eigenvalues in this subset of fixed $r=q\log \rho$ is now to become very small, of order $1/\rho$.

In the above discussion,  we see that generally the exponent $\energy_1(r)$ is not only continuous, but also differentiable in $r$. This is in disagreement to the prediction by \cite{Zheng2003_DiversityMultiplexing, Azarian2007_finite_rate_DMT} that when $\rho\rightarrow \infty$, the outage has a piecewise linear behavior. The length of these segments is $\Delta R\approx\log\rho$, or $\Delta r\approx\log\rho/N$. Thus for these segments to be pronounced we need
\begin{equation}\label{eq:LD_validity}
N \ll \log\rho
\end{equation}
for large $\rho$. This provides a limit on the formal limitations of our large deviations (LD) approach. In particular, clearly the antenna number $N$ has to be large, as in the Gaussian case. But, in contrast to the Gaussian approximation, there is no constraint here that the deviation of the rate from the ergodic rate has to be small, as in (\ref{eq:Gaussian_validity}). Thus the {\it scale } of $N$ at which the method should break down is given by $\log\rho$ for large $\rho$, rather than $\rho$ itself. This is corroborated in the numerical results in the next section. Surprisingly, however, the analysis in this section shows that the {\it form} of the DMT exponent (\ref{eq:DMT_def}) is correctly predicted within the LD approach in (\ref{eq:E1-E0_dmt}).

\subsection{Large $\rho$ Approximation: $r>r_{\text{erg}}$}
\label{sec:DMT_region_r>rerg}

The regime of large $\rho$ and fixed $q=r/\log\rho\leq 1$ is relevant in the analysis of the link-level outage probability. However, the opposite regime of $q>1$ is also of interest in a cellular setting with many multi-antenna users receiving data in a TDMA fashion from a single multi-antenna base-station.\footnote{In that case a MAC-layer scheduler would be transmitting to the user with the best channel, for example.} In this context, to analyze the system level throughput, it is the higher end of the probability distribution of the link-level mutual information that is important \cite{Hochwald2002_MultiAntennaChannelHardening, Bender2000_HDR_ComMagReview}. Therefore, it is worthwhile to calculate the probability distribution of $r$ for large $\rho$ with $q>1$.

Interestingly enough, the behavior here is quite different from the $q<1$ case. Here $k\sim \rho^{q-1}$ and
\begin{eqnarray}
\label{eq:a_q>1_regime}
  a &\sim& \left(\sqrt{k+\beta}-1\right)^2 \\
\label{eq:b_q>1_regime}
  b &\sim& \left(\sqrt{k+\beta}+1\right)^2
\end{eqnarray}
resulting to
\begin{equation}\label{eq:E1-E0_q>1}
    \energy_1-\energy_0 \sim \rho^{q-1} = \frac{e^r}{\rho}
\end{equation}
independent of $\beta$. The resulting probability distribution of $r$ is
\begin{equation}\label{eq:P(r)_r>>r_erg}
    P(r) \sim e^{-N^2 e^r/\rho}
\end{equation}
We see that when $N$ is not too small, the probability of finding $I_N$ significantly larger than its ergodic value is extremely small (in fact, doubly exponentially small in $r$). This is the manifestation of the fact that scheduling the best user in a MAC-layer in a multi-antenna setting does not seem to provide any clear advantage. Interestingly, in \cite{Hochwald2002_MultiAntennaChannelHardening} the authors have the same conclusion, even though they assume a Gaussian distribution for $I_N$ even for its tails. Here we see that the distribution of $I_N$ goes to zero for $r>r_{erg}$ in a rate even faster than Gaussian, thereby making the above conclusion, to which they also reached even stronger.

This result has the following intuitive explanation. For large $\rho$ and $r>r_{erg}$ all eigenvalues of the matrix $\bH^\dagger\bH$ will be large and the only constraint imposed upon them is (\ref{eq:mutual_information_condition}). Thus, we may say that all of them are constrained by the condition $r\sim \log(1+\rho \lambda_i)\sim\log\rho\lambda_i$ i.e. $\lambda_i\sim e^r/\rho$. In this limit, the exponent is roughly $N$ times the sum of the eigenvalues.

\subsection{Limit $r\rightarrow 0$}
\label{sec:r->0}

The final regime that is interesting to analyze is when  $r\rightarrow 0$, independently of $\rho$.  In this regime the solution of (\ref{eq:mut_info_beta>1}) (\ref{eq:mut_info_beta=1_r<r_c}) for small $r$ is $r\sim \beta/|k|$ for $k\rightarrow-\infty$ and the corresponding values of $a,b$ are:
\begin{eqnarray}
\label{eq:a_r->0_regime}
  a &\sim& \frac{r}{\rho \beta}\left(\sqrt{\beta}-1\right)^2  \\
\label{eq:b_r->0_regime}
  b &\sim& \frac{r}{\rho \beta}\left(\sqrt{\beta}+1\right)^2
\end{eqnarray}
resulting in:
\begin{equation}\label{eq:E1-E0_r->0}
    \energy_1-\energy_0 \sim -\beta \log \left[\frac{e r}{\beta \rho}\right].
\end{equation}
where $e$ is the Euler number. This means that the probability distribution $P_N(r)$ has a tail of the form
\begin{equation}
\label{eq:P(r)_r->0}
    P(r) \sim \left(\frac{r e}{\rho \beta}\right)^{MN}
\end{equation}
The above behavior of $P_N(r)$ for small $r$ is easy to understand: for $r$ to be small, we need all matrix elements of the matrix $\bH$ to be small. In fact, since $\bH$ appears in a quadratic way in the mutual information equation (\ref{eq:logdet_def}) we need all $MN$ elements of $\bH$ to be less than $O(\sqrt{r/\rho})$. However, there are $2MN$ real degrees of freedom in the $M\times N$ complex matrix $\bH$. Hence the allowed volume of space scales as $(r/\rho)^{MN}$ as above.

It should also be noted that the behavior $P(r)\sim \rho^{-MN}$ of the mutual information cumulative distribution function is precisely what is known as the ``full diversity'' of error probability, i.e., the SNR exponent of error probability for fixed but very small rate $R$ while SNR $\rho$ increases is $\rho^{-MN}$, which corresponds to the left extreme point of the Zheng-Tse exponent\cite{Zheng2003_DiversityMultiplexing}.

\section{Numerical Simulations}
\label{sec:numerical simulations}

To test the applicability of this approach, we have performed a series of numerical simulations and have compared our large deviations (LD) approach to other popular approximations.

\begin{figure*}
\centerline{%
\subfigure[$\rho=-10dB$]
{\epsfxsize=.33\textwidth
 \epsffile{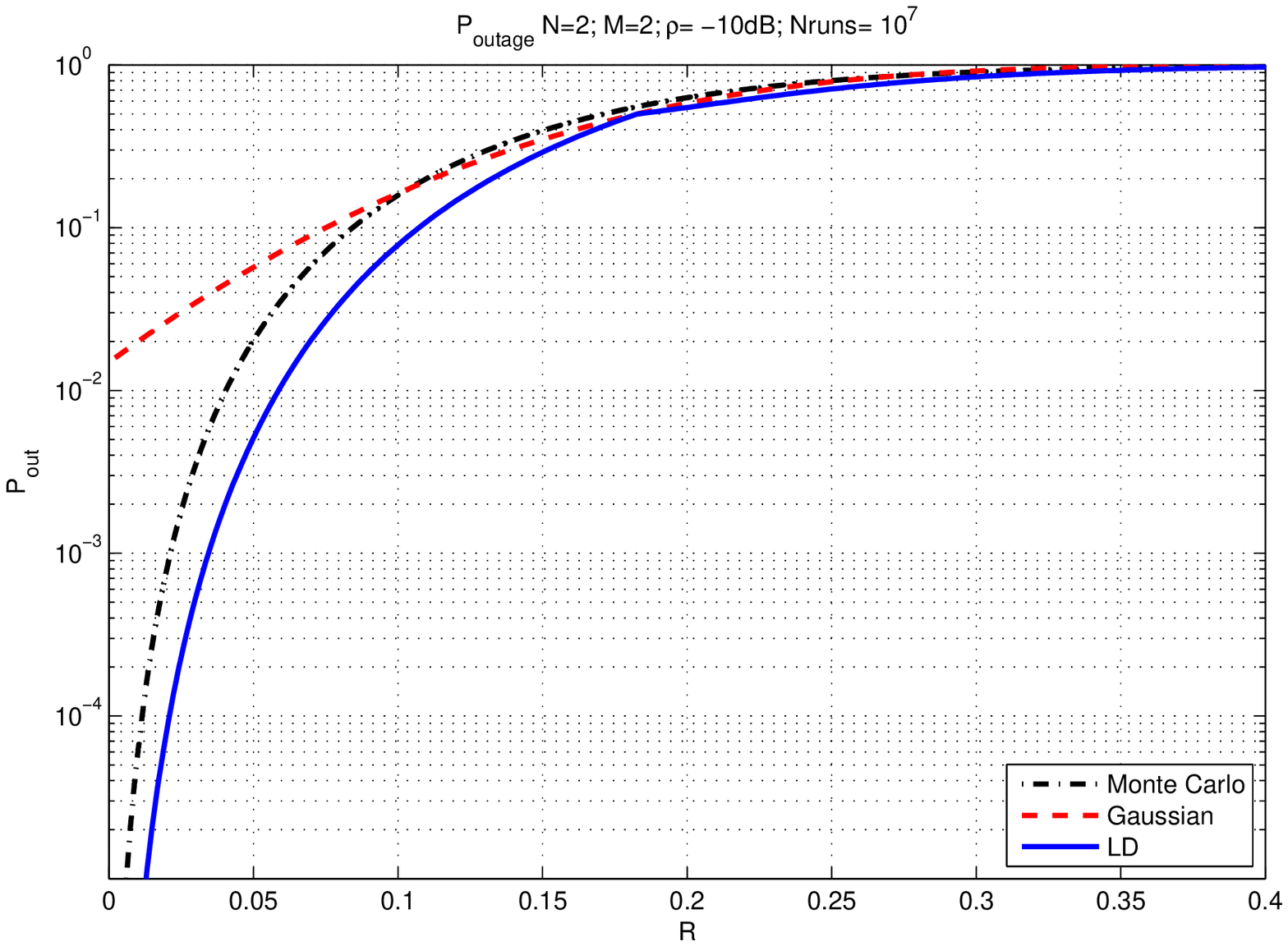} }
\hfil %
\subfigure[$\rho=0dB$]
{\epsfxsize=.33\textwidth
 \epsffile{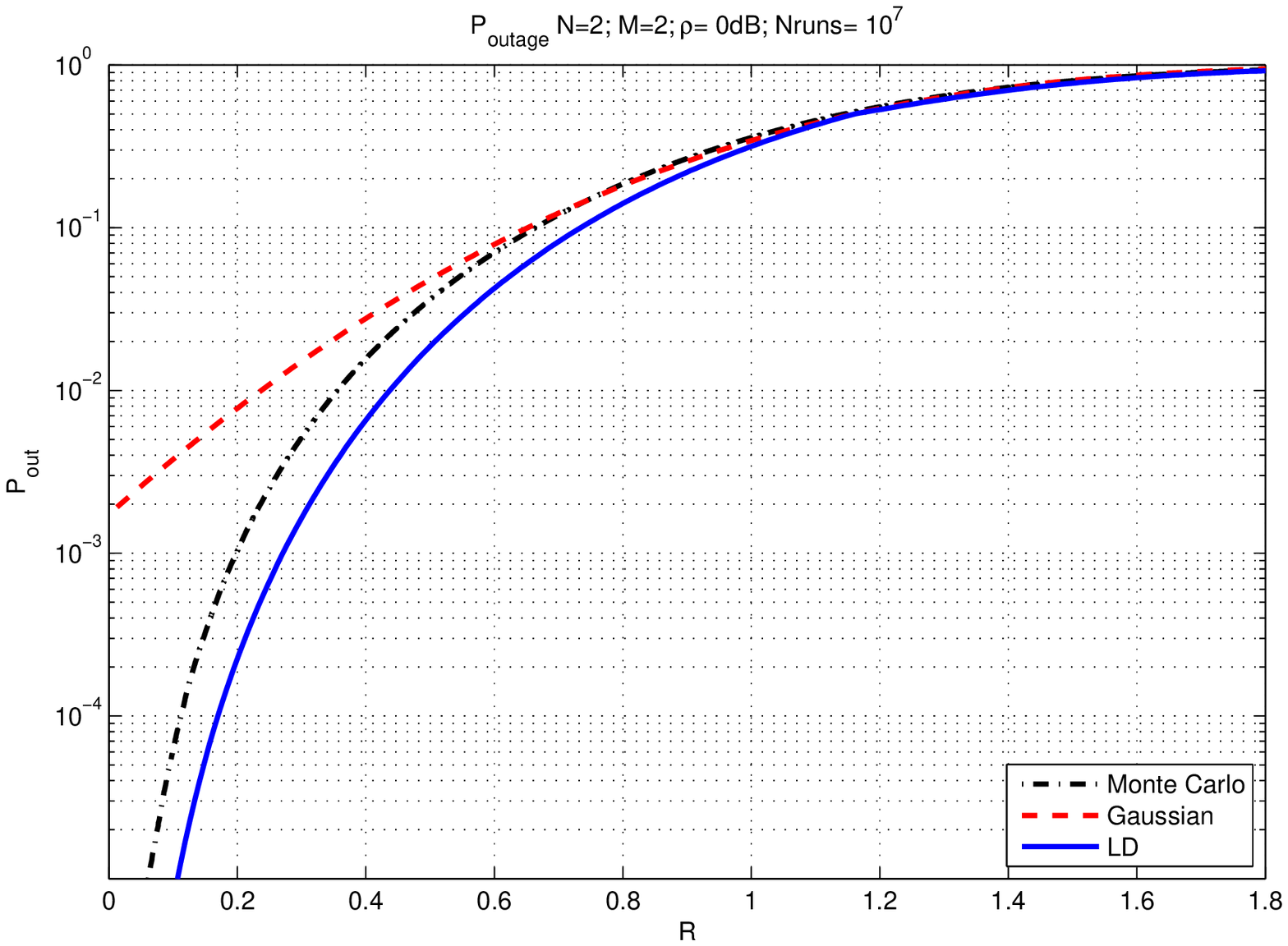}}
\hfil %
\subfigure[$\rho=10dB$]
{\epsfxsize=.33\textwidth
 \epsffile{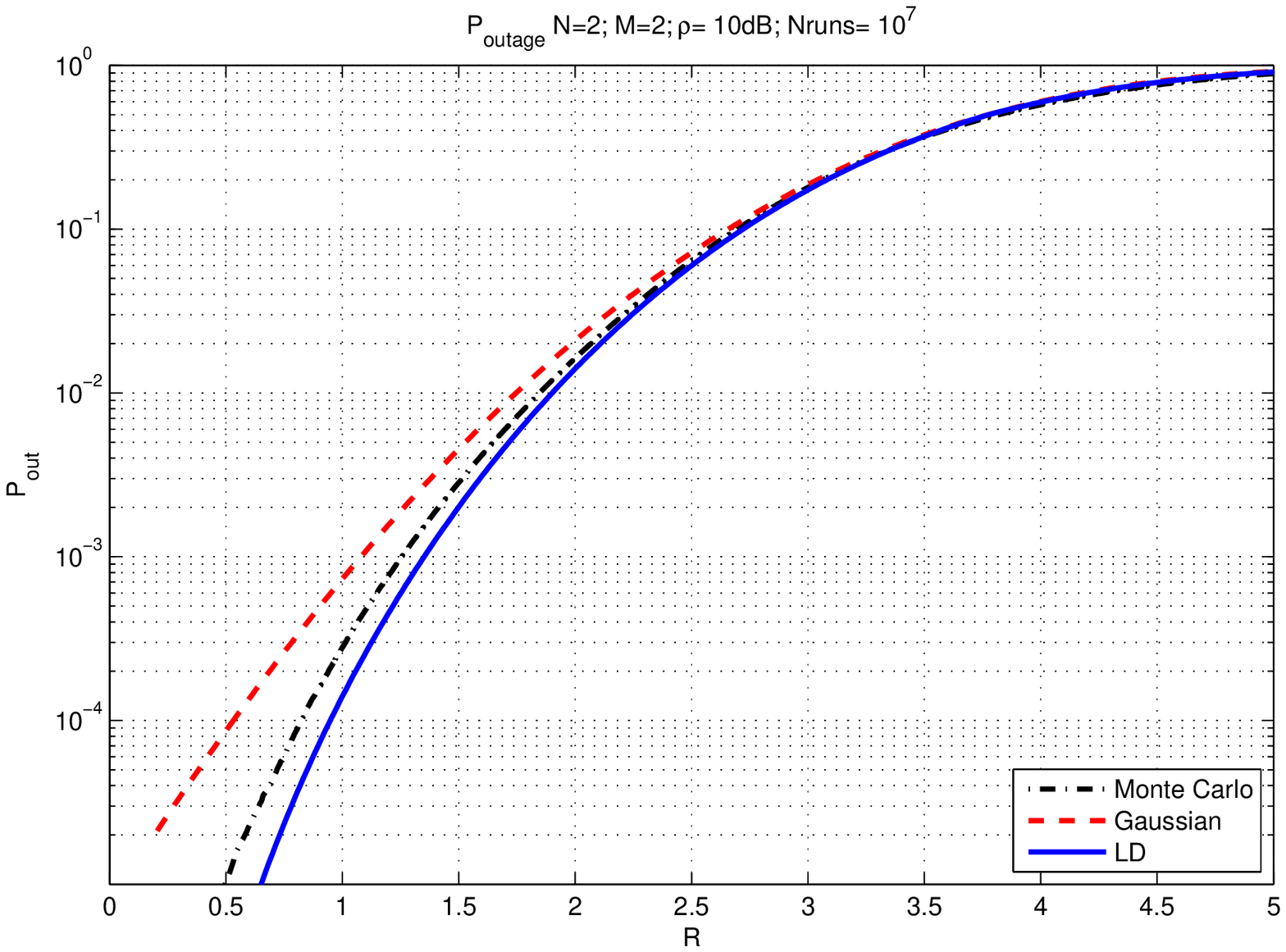} }  }
\caption{Comparison of the outage probability curves for $N=M=2$  of the Large Deviation result with the Gaussian approach and Monte-Carlo simulations. The three subplots are for different SNR values:  (a) with $\rho=-10dB$, (b) with $\rho=0dB$ and (c) with $\rho=10dB$. We see that for decreasing $\rho$ the discrepancy between the Gaussian curve (dashed) and the other two, i.e.  LD (solid)) and simulated (dash-dotted) is increasing.} \label{fig:CDF_N2M2}
\end{figure*}

\begin{figure*}
\centerline{%
\subfigure[$\rho=-10dB$]
{\epsfxsize=.33\textwidth
 \epsffile{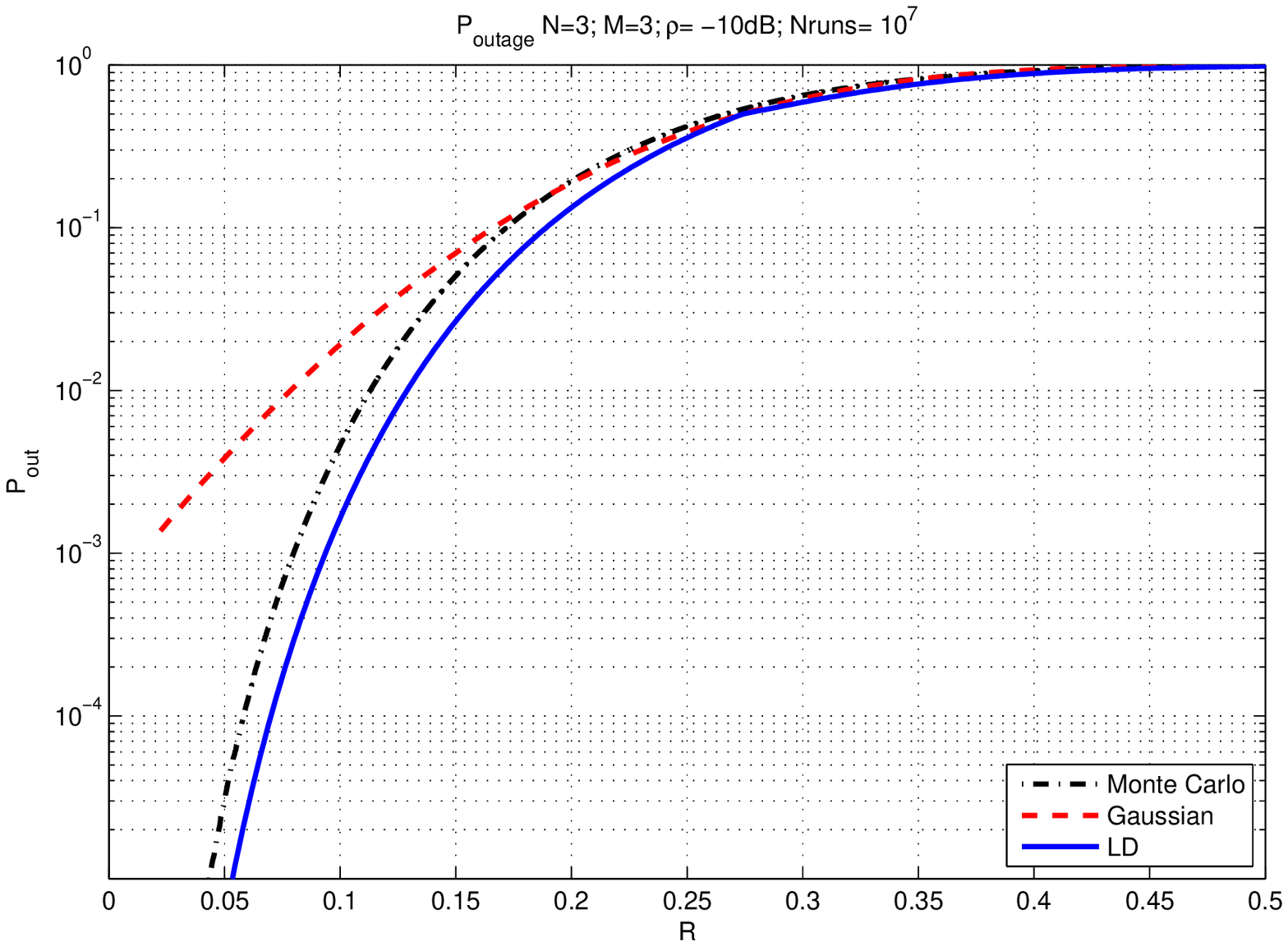} }
\hfil %
\subfigure[$\rho=0dB$]
{\epsfxsize=.33\textwidth
 \epsffile{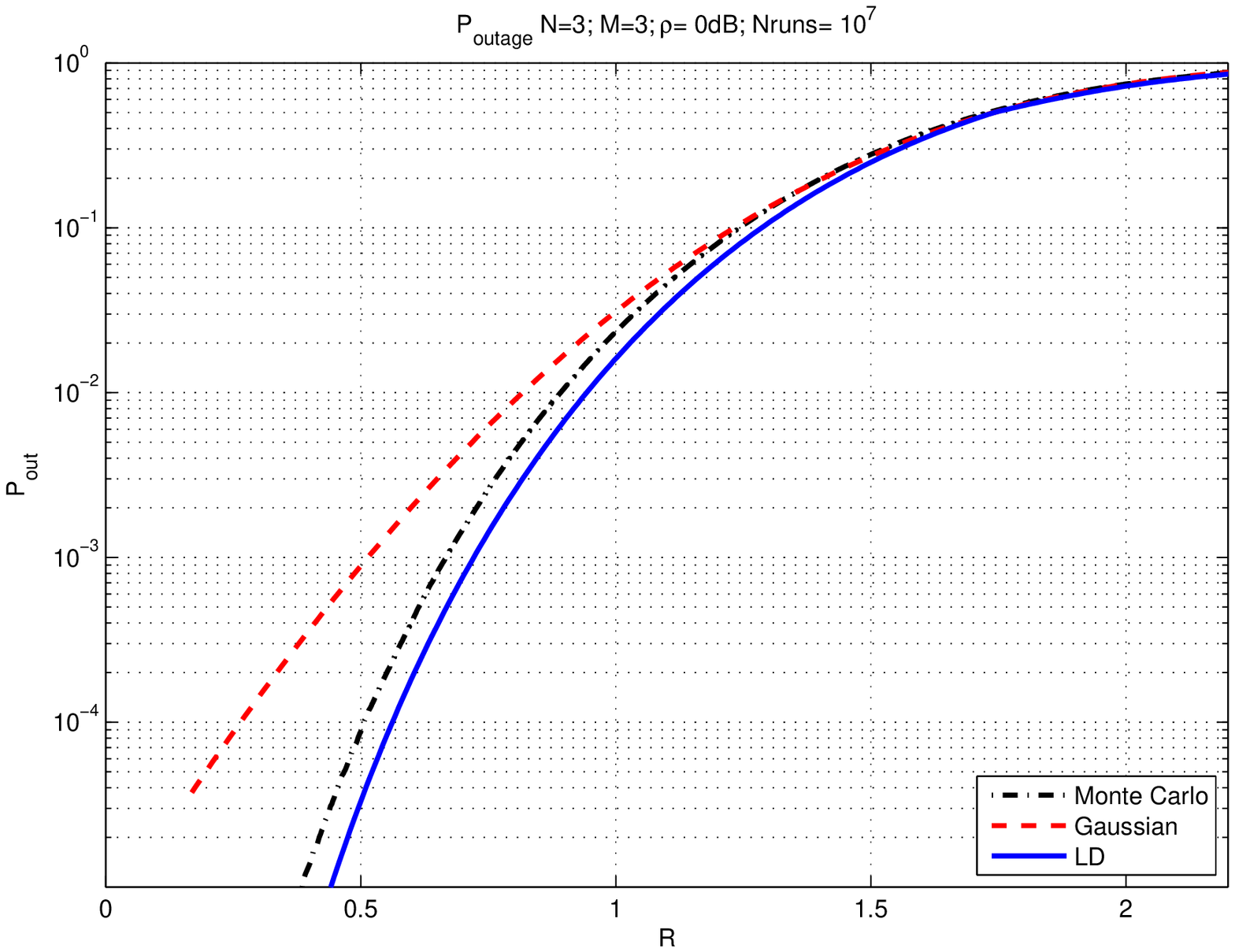}}
\hfil %
\subfigure[$\rho=10dB$]
{\epsfxsize=.33\textwidth
 \epsffile{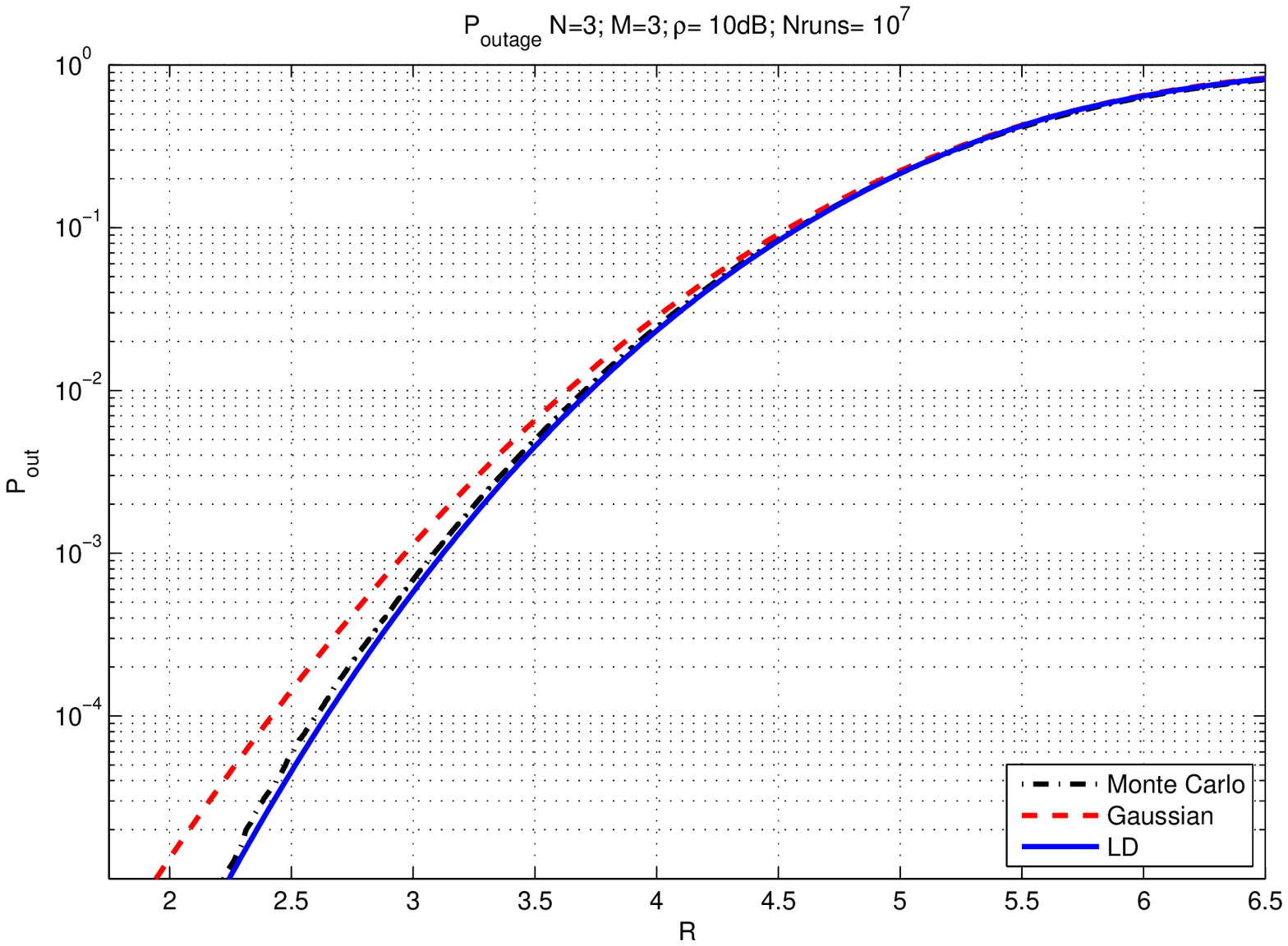}}}
\caption{Comparison of the outage probability curves for $N=M=3$  of the Large Deviation result with the Gaussian approach and Monte-Carlo simulations. The three subplots are for different SNR values:  (a) with $\rho=-10dB$, (b) with $\rho=0dB$ and (c) with $\rho=10dB$. We see that for decreasing $\rho$ the discrepancy between the Gaussian curve (dashed) and the other two, i.e.  LD (solid)) and numerical (dash-dotted) is increasing. Comparing the $N=3$ with the $N=2$ results, we see that the former are generally closer to the simulated curve, nevertheless, the Gaussian curve is always clearly further away.} \label{fig:CDF_N3M3}
\end{figure*}

We start with the case of small rates $r$. In this limit the Gaussian approximation is guaranteed to give misleading results. For example, the Gaussian approximation predicts a finite outage probability at zero rate, while this is clearly wrong. The LD approximation, on the other hand, correctly predicts that the outage probability goes to zero at small $r$, as seen in (\ref{eq:P(r)_r->0}). In Figs. \ref{fig:CDF_N2M2} and \ref{fig:CDF_N3M3} we plot the outage probability of the LD approach with the Gaussian and Monte Carlo simulations for low rates, small $\rho$ and small square ($2\times 2$ and $3\times 3$) antenna arrays. The comparison shows that while the Gaussian curves miss the correct outage, the LD curves remain close to the simulated ones, even for the $2\times 2$ MIMO system. It is worthwhile to mention that the Gaussian outage probability is consistently greater than the correct (simulated) one. The reason for this can be traced to the fact that for all $\beta=1$ and all values of $\rho$, the third derivative of the exponent ${\cal E}_1(r)-{\cal E}_0$ with respect to $r$ evaluated at $r_{erg}$, i.e. $s_{erg}(\rho)$ in (\ref{eq:E1-E0_small_k2}) is negative. Disturbing away from the peaks of the distribution we have
\begin{equation}\label{eq:P_out_gaussian_approx}
\log P_{out,Gaussian}(r)\approx -\frac{N^2(r-r_{erg})^2}{2v_{erg}}
\end{equation}
while
\begin{equation}\label{eq:P_out_gaussian+3rd_approx}
\log P_{out}(r)\approx -\frac{N^2(r-r_{erg})^2}{2v_{erg}} - \frac{s_{erg}N^2(r-r_{erg})^3}{6}
\end{equation}
We may thus conclude that when $r<r_{erg}$ and $s_{erg}<0$ we should have $P_{out,Gaussian}>P_{out}$. From Fig. \ref{fig:E3_vs_rho} we see that for increasing $\rho$, $s_{erg}$ decreases in absolute size, which correctly predicts that the discrepancy between the Gaussian and the Monte-Carlo curves (and LD) decreases for larger $\rho$.

\begin{figure*}
\centerline{%
\subfigure[Complementary CDF for $\rho = 20dB$]
{\epsfxsize=.49\textwidth
 \epsffile{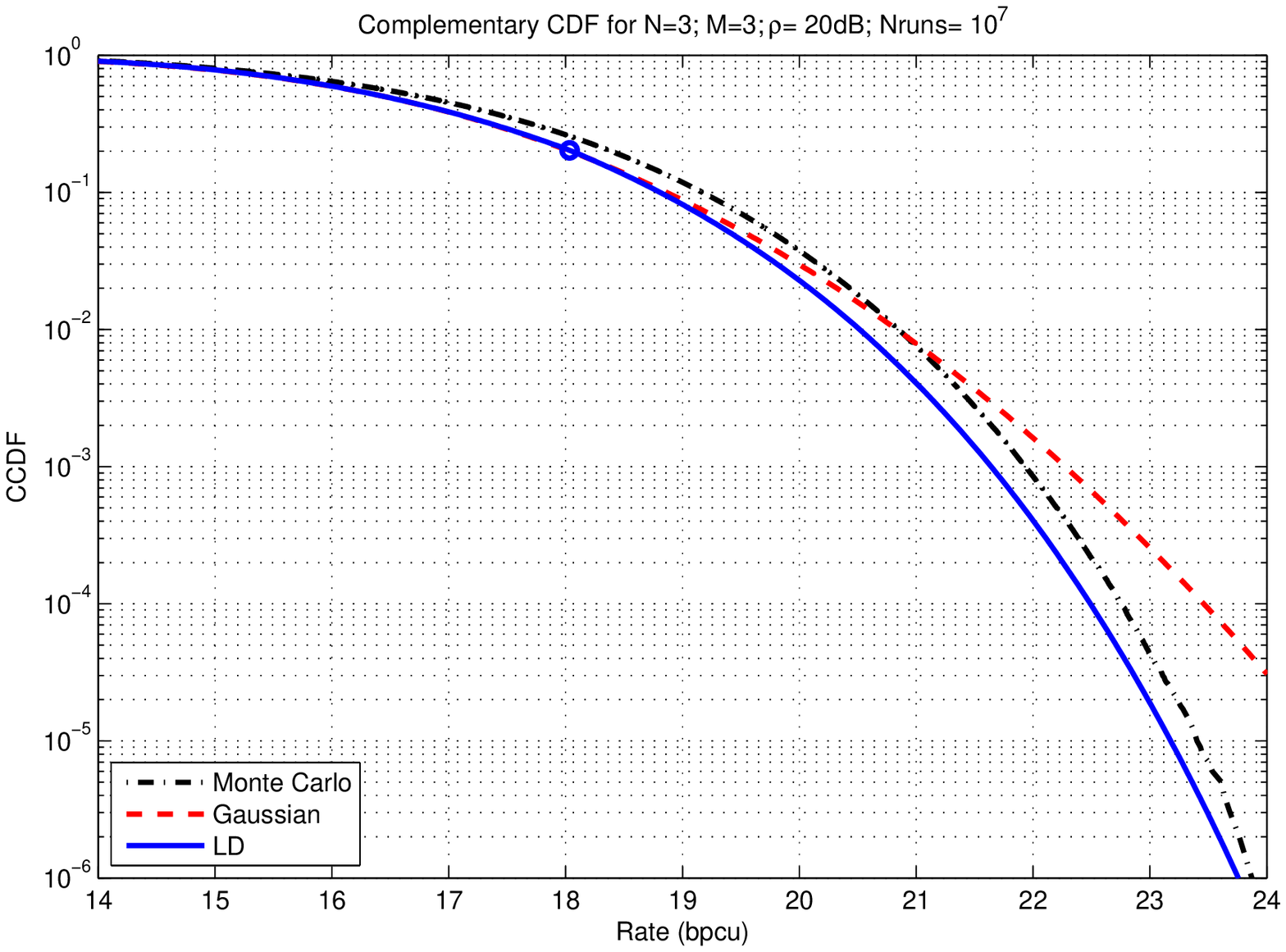} }
\hfil %
\subfigure[Complementary CDF for $\rho = 50dB$]
{\epsfxsize=.49\textwidth
 \epsffile{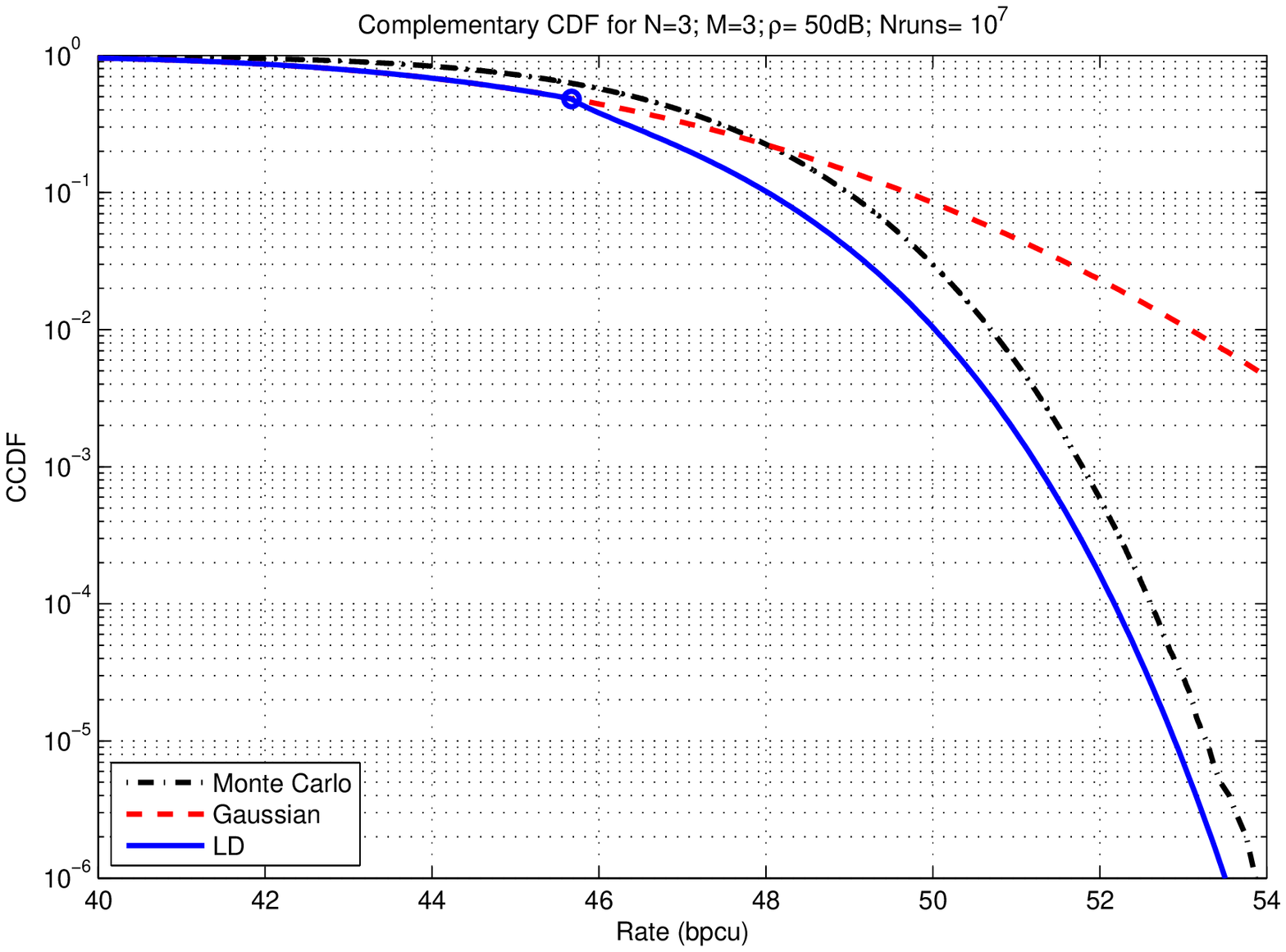}}}
\caption{In these figures we depict the complementary cumulative distribution function (CCDF) of the mutual information for the antenna array $3\times 3$. In this region of parameters we compare the the current methodology (LD) (solid) with numerical Monte-Carlo simulations (dash-dotted) and  the Gaussian approximation (dashed). We also depict the rate value $r_{c}$ at which, for the given SNR, the exponent dependence on $r$ changes from (\ref{eq:E1_beta=1_r<rc}) to (\ref{eq:E1_beta=1_r>rc}). We see that at that point the distribution starts deviating strongly from the Gaussian approximation. It should be pointed out that this point corresponds to a mild {\it phase transition} as discussed in Section \ref{sec:case r<r_c} and also analyzed in a different context in \cite{Vivo2008_DistributionsConductanceShotNoise}. Nevertheless, in both moderate and large SNRs the LD curve is consistently  close to the simulated curves. (a) CCDF for $\rho=20dB$ (b) CCDF for $\rho=50dB$} \label{fig:CCDF_N3M3}
\end{figure*}

We have also analyzed the probability distribution for rates greater than the ergodic rate $r>r_{erg}$. Even though this region is not relevant for the outage probability evaluation, it is important in the analysis of the multiuser capacity for MIMO links in a multi-user setting with a greedy scheduler, such as a maximum rate scheduler.\cite{Hochwald2002_MultiAntennaChannelHardening} In such a case, the multiuser diversity gain comes from the opportunity the scheduler has to schedule transmission to users when their fading rates are greater than their mean. Thus it is important to understand the tails of the distribution in this region. In Fig. \ref{fig:CCDF_N3M3} we obtained the complementary CDF (CCDF) of the mutual information, i.e. $1-P_{out}(r)$, for a $3\times 3$ setting. Here the probability of finding users with high rates falls faster than the Gaussian, especially in Fig. \ref{fig:CCDF_N3M3}b for large $\rho$. We also find that the LD approximation follows the Monte Carlo simulations more accurately than the Gaussian curve, especially for lower outages. In this situation it is worth pointing out that the argument mentioned above regarding the sign of $s_{erg}$ would make $1-P_{out}(r)$ smaller in the Gaussian approximation compared to the correct result. We see that this only occurs for rates relatively close to the peak. In contrast, for rates greater than the critical rate $r_c(\rho)$ the behavior  of the numerical and the LD outage probability changes markedly and they both become substantially smaller than the Gaussian curve. This is not surprising in view of the phase transition occurring at $r=r_c(\rho)$ as discussed in Section \ref{sec:case r<r_c}.

We next analyzed the outage probability as a function of the SNR. The outage has been analyzed in the large SNR limit for finite rates in \cite{Azarian2007_finite_rate_DMT}, where they have dubbed this analysis as throughput reliability tradeoff (TRT). This model provides a piecewise linear function of the outage probability, which for completeness is provided below:
\begin{eqnarray}\label{eq:TRT_model}
    \log_2 P_{out} &\approx& c(k) R - g(k) \log_2\rho \\ \nonumber
    c(k) &=& M+N-2k-1 \\ \nonumber
    g(k) &=& MN -k(k+1)
\end{eqnarray}
when $\rho$ is large and $k\log_2\rho <R<(k+1)\log_2\rho$.

\begin{figure*}
\centerline{%
\subfigure[Outage for $N=3$, $M=6$ and R=4, 16, 28, 40, 52 bpcu ]
{\epsfxsize=.49\textwidth
 \epsffile{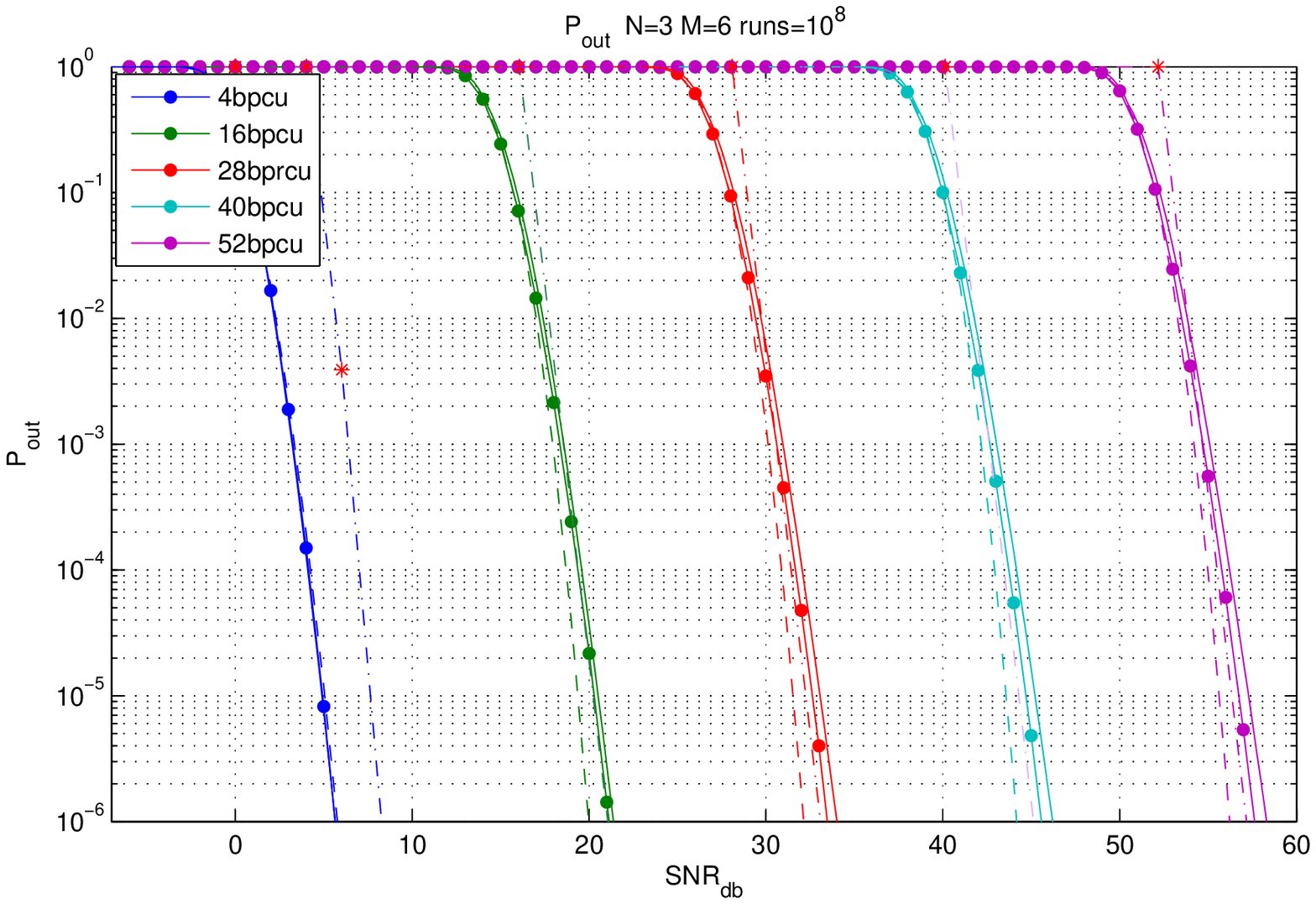} }
\hfil %
\subfigure[Outage probability only for R=16 bpcu ]
{\epsfxsize=.49\textwidth
 \epsffile{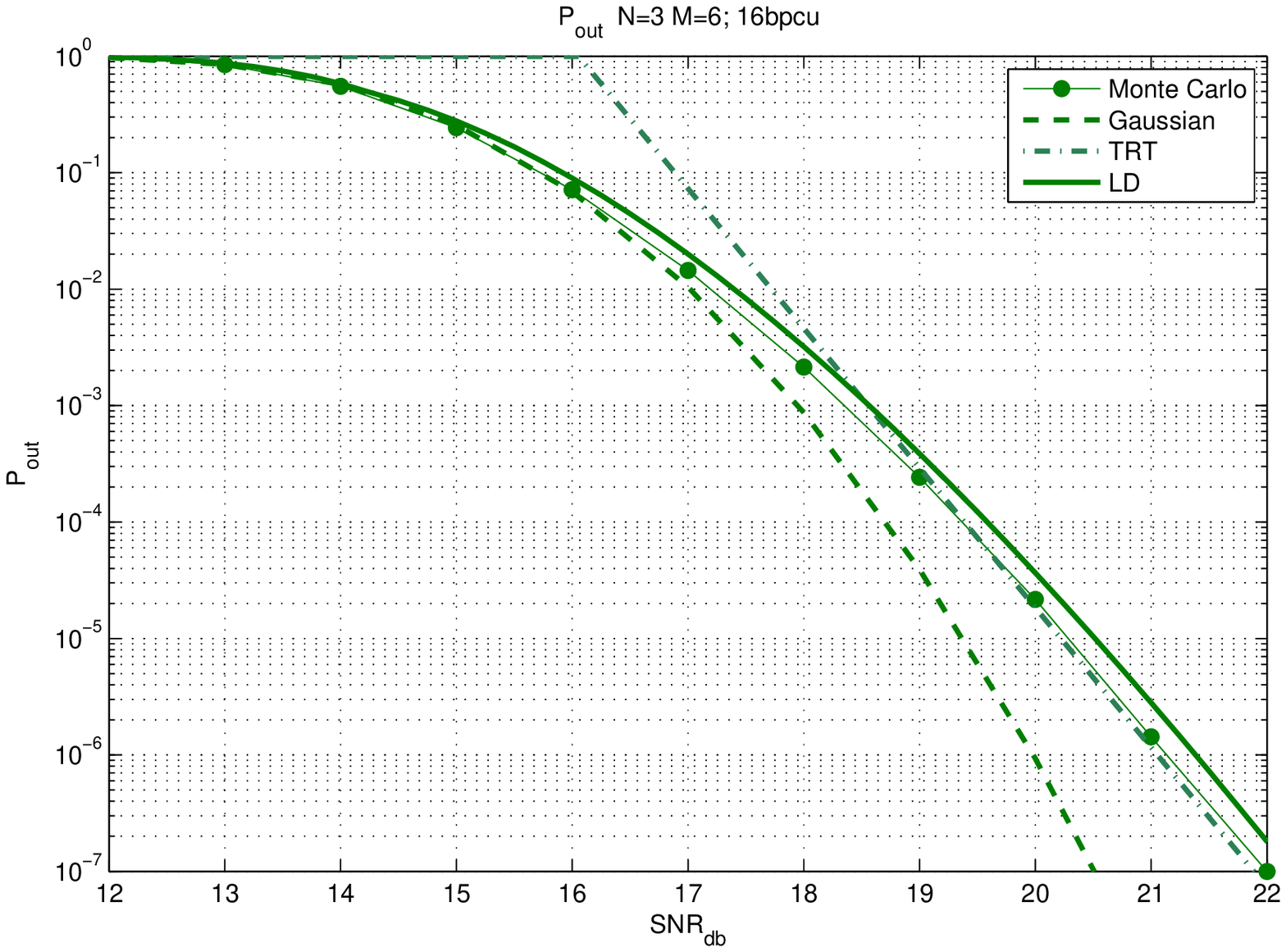}}}
\caption{In these figures we depict the outage probability as a function of SNR for the antenna array $3\times 6$. The current methodology (LD) (solid) is compared with numerical Monte-Carlo simulations ($10^8$ runs, solid with dots) and two other approximations, the Gaussian (dashed) and the Throughput-Reliability-Tradeoff (TRT) approximation (dash-dot), analyzed in \cite{Azarian2007_finite_rate_DMT}. The red stars on the TRT curve depict the points at which the lines change slope. (a) In this figure, we collectively plot  the curves at a number of bpcu values. At this scale all three candidates behave rather well, except perhaps for the TRT curve at the lowest bpcu value (R=4). (b) Nevertheless, zooming in for the R=16 bpcu case, we see that both the TRT and Gaussian approximations significantly depart from the numerical curve, at low and high SNRs correspondingly. In contrast, the LD curve is consistently closer to the numerics.} \label{fig:Pout_N3M6}
\end{figure*}

\begin{figure*}
\centerline{%
\subfigure[Outage for $N=M=3$, and R =4, 16, 28, 40, 52 bpcu ]
{\epsfxsize=.49\textwidth
 \epsffile{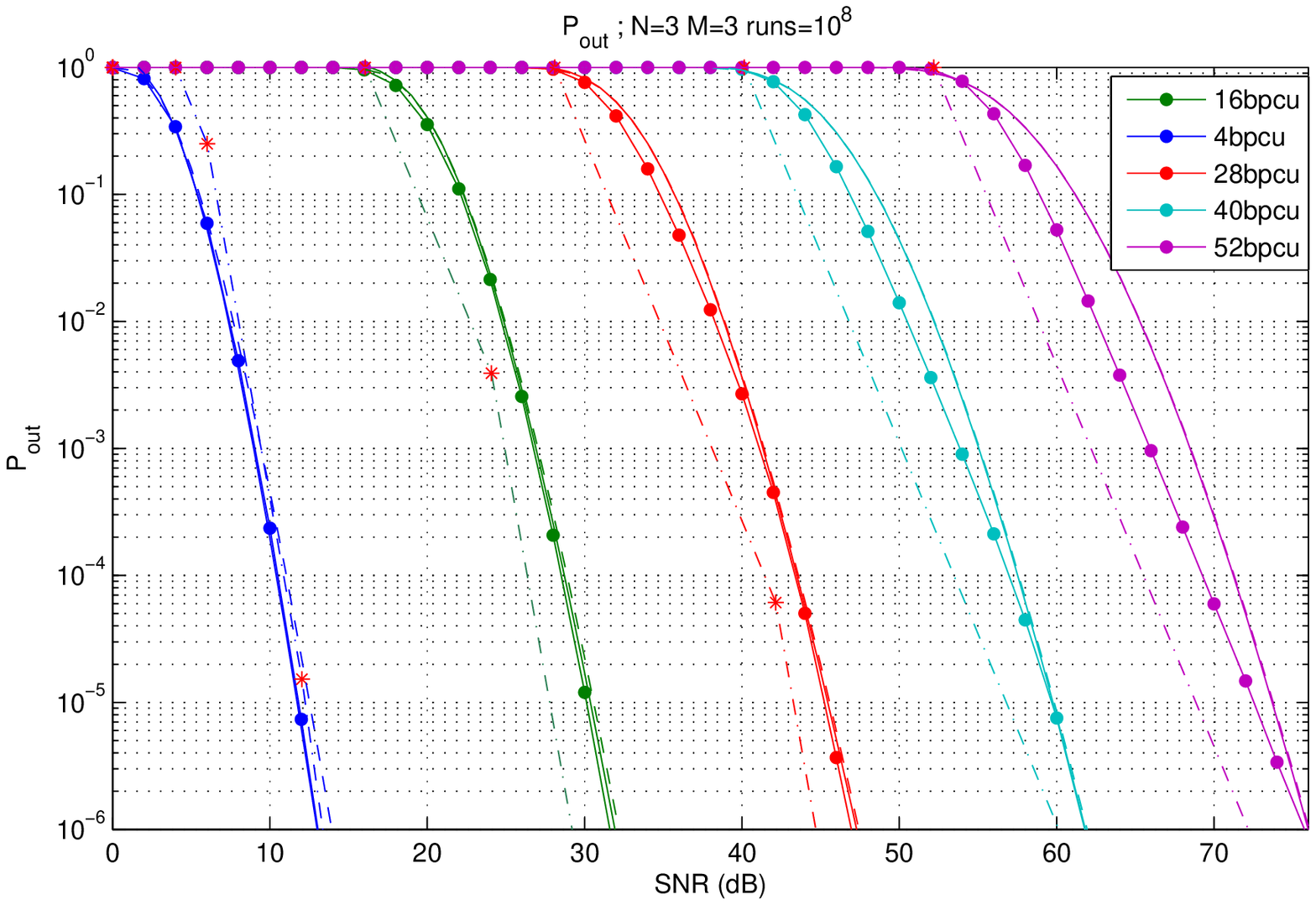} }
\hfil %
\subfigure[Outage for $N=M=6$, and R =4, 16, 28, 40, 52 bpcu ]
{\epsfxsize=.49\textwidth
 \epsffile{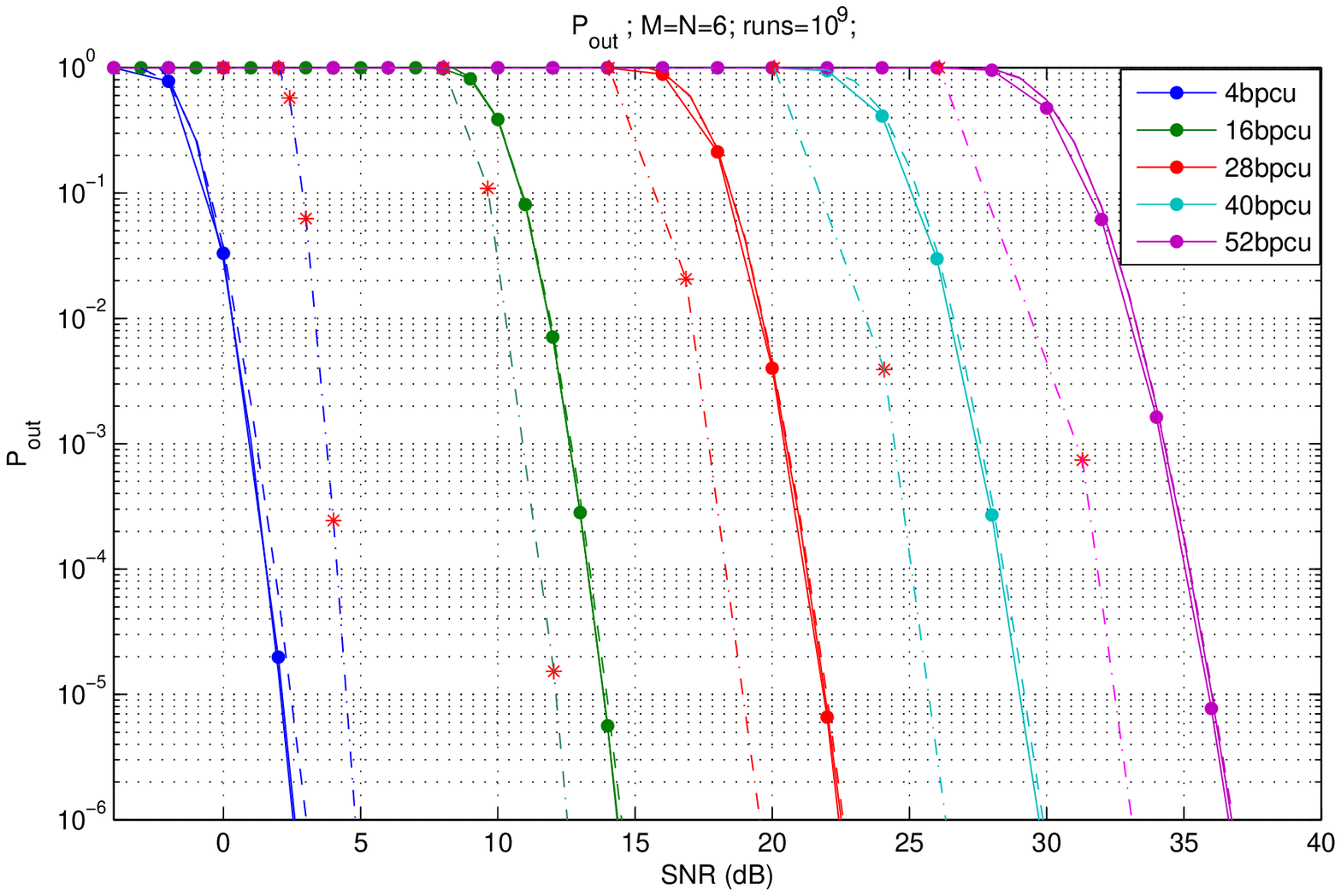}}}
\caption{In these figures we depict the outage probability as a function of SNR for the antenna arrays $N=M$. The current methodology (LD) (solid) is compared with numerical Monte-Carlo simulations (solid with dots) and two other approximation, the Gaussian (dashed) and the Throughput-Reliability-Tradeoff (TRT) approximation (dash-dot), analyzed in \cite{Azarian2007_finite_rate_DMT}. The red stars on the TRT curve depict the points at which the lines change slope.
(a) Curves for outage probability versus SNR for the antenna array $3\times 3$ for the same bpcu values as in Fig. \ref{fig:Pout_N3M6}. In contrast to that figure, for very large values of SNR ($\rho>45dB$) both the LD and Gaussian approximations deviate from the numerics ($10^8$ runs), which exhibits a  linear behavior (in a log-log plot). This deviation of the LD approximation is expected. Here the number of antennas is still quite small ($N=3$), while the SNR is extremely large, making the LD approximation (in addition to the Gaussian) not valid. In these extreme SNRs the TRT approximation seems to have the correct slope, but also misses the exact value of the outage probability. For more reasonable SNR, the LD is quite close to the numerical plot. (b) Curves for outage probability versus SNR for the antenna array $6\times 6$. In this case, the LD approximation works well even for such large SNRs.} \label{fig:Pout_N3M3}
\end{figure*}

This piecewise linear behavior however is observable only at extremely high rates and SNRs, which may not necessarily be relevant for realistic MIMO systems. We analyzed the case of $3\times 6$, $3\times 3$ and $6\times 6$ arrays in Figs. \ref{fig:Pout_N3M6} and \ref{fig:Pout_N3M3}. In all three we have found that the LD approximation agrees with simulations over a wide region of rates $r$ and SNR $\rho$. Characteristic is Fig. \ref{fig:Pout_N3M6}b, where the TRT curve is accurate in large SNR, the Gaussian is accurate in low SNR, but the LD curve is consistently closer to the correct outage. For the $N=M=3$ case and extremely high SNRs and rates the piecewise linear behavior predicted by TRT starts becoming visible. Nevertheless, even in those high rates the TRT curve also fails to give quantitatively correct outage estimates and the LD curve is still closer to the correct outage.

It is sensible to point out that here the Gaussian outage probability is consistently less than the  simulated and the LD values. In this case the argument made above for $s_{erg}$ is reversed. As can be seen in Fig. \ref{fig:E3_vs_rho} for $\beta=2$ and large $\rho$ the sign of $s_{erg}$ is opposite, i.e. we have $s_{erg}>0$ and hence indeed we should have $P_{out,Gaussian}<P_{out}$.

%
%
%

\begin{figure}[htb]
\centerline{\epsfxsize=1.0\columnwidth\epsffile{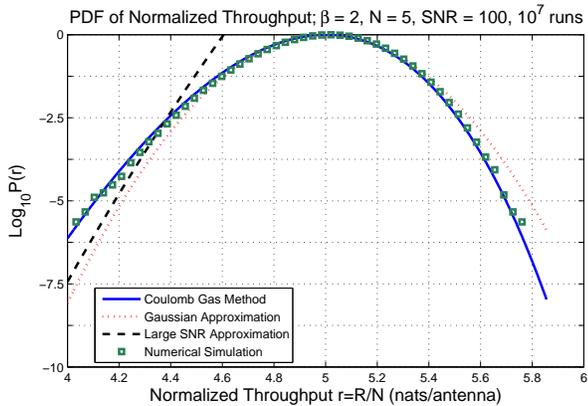}}
\caption{Plot of the logarithm of the normalized probability distribution curve of the mutual information $I_N/N$ for $\beta=2$ and comparison to the Gaussian approximation and the large-$\rho$ asymptotic result obtained by (\ref{eq:DMT_def}) \cite{Zheng2003_DiversityMultiplexing}. The numerical result for $N=5$ follows closely our result, even at large $\rho=100$.} \label{fig:E1_N5_beta2_SNR_100}
\end{figure}

In Fig. \ref{fig:E1_N5_beta2_SNR_100}, we plot the logarithm of the appropriately normalized probability density function (PDF) $P_N(r)$ as a function of the throughput $r$ and we compare the result with the two other asymptotic forms, namely the Gaussian approximation of the mutual information \cite{Moustakas2003_MIMO1} and the large-$\rho$ asymptotic result given by (\ref{eq:DMT_def}) \cite{Zheng2003_DiversityMultiplexing}. We see that our result performs much better at low outage, even at moderately large $\rho =20dB$.

As discussed in the Introduction, the LD method is the correct generalization of the  Gaussian approximation to capture the tails of the distribution of the mutual information. As a result, it is expected to give increasingly accurate results as the antenna number $N$ increases. In the above comparisons we have compared the LD method with numerical simulations focusing on its tails (low outage $P_{out}$ or low values of $1-P_{out}$) for small antenna numbers. We have found that the LD approximation behaves well even at these values of $N$. The discrepancy between the LD approximation and Monte Carlo simulations becomes smaller for larger $N$ as seen in Fig. \ref{fig:E1_N5_beta2_SNR_100}.

\begin{figure*}
\centerline{%
\subfigure[PDF $N=2$, $M=4$ for $\rho=20dB$]
{\epsfxsize=.49\textwidth
 \epsffile{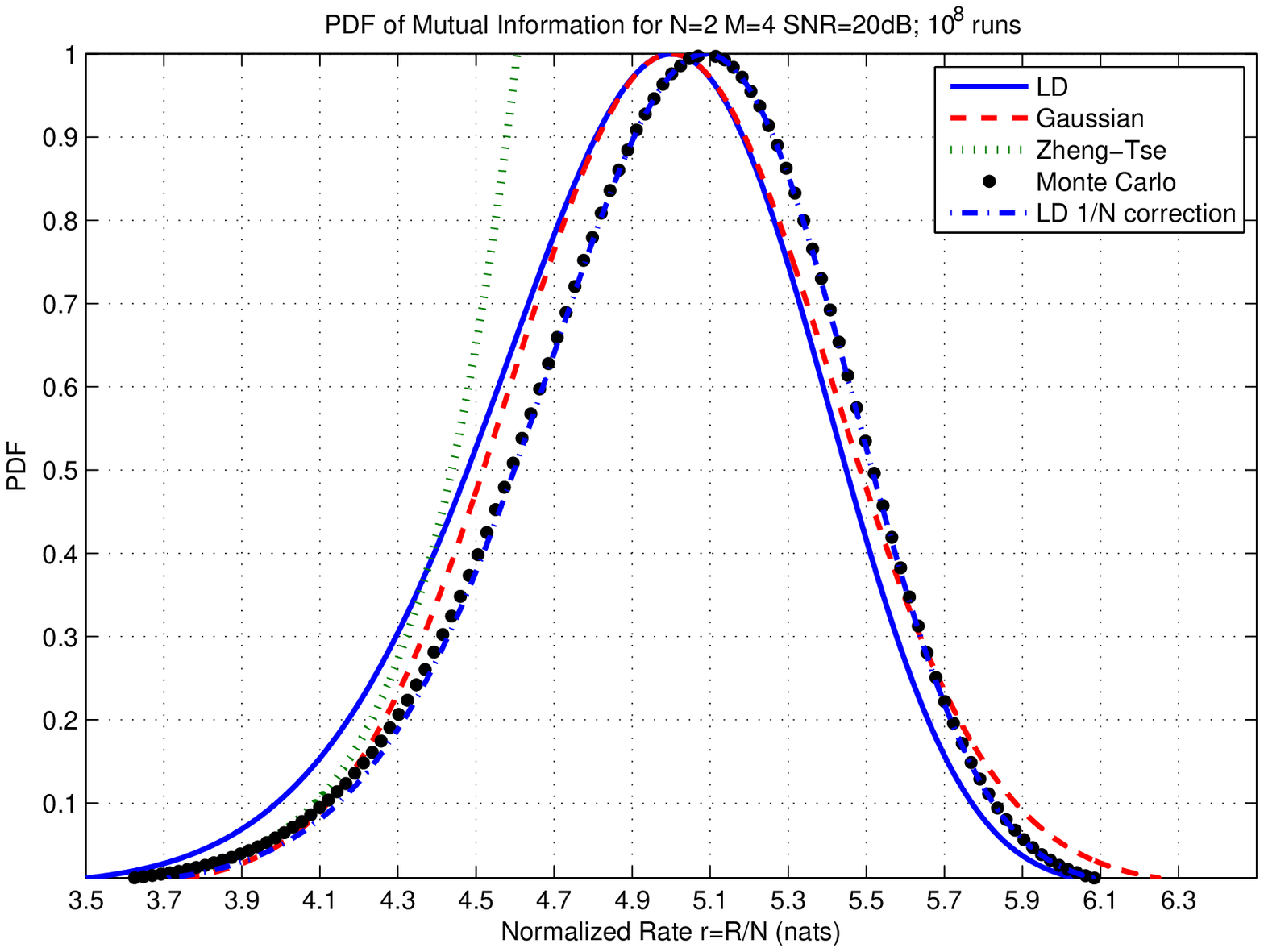} }
\hfil %
\subfigure[PDF $N=2$, $M=4$ for $\rho=50dB$]
{\epsfxsize=.49\textwidth
 \epsffile{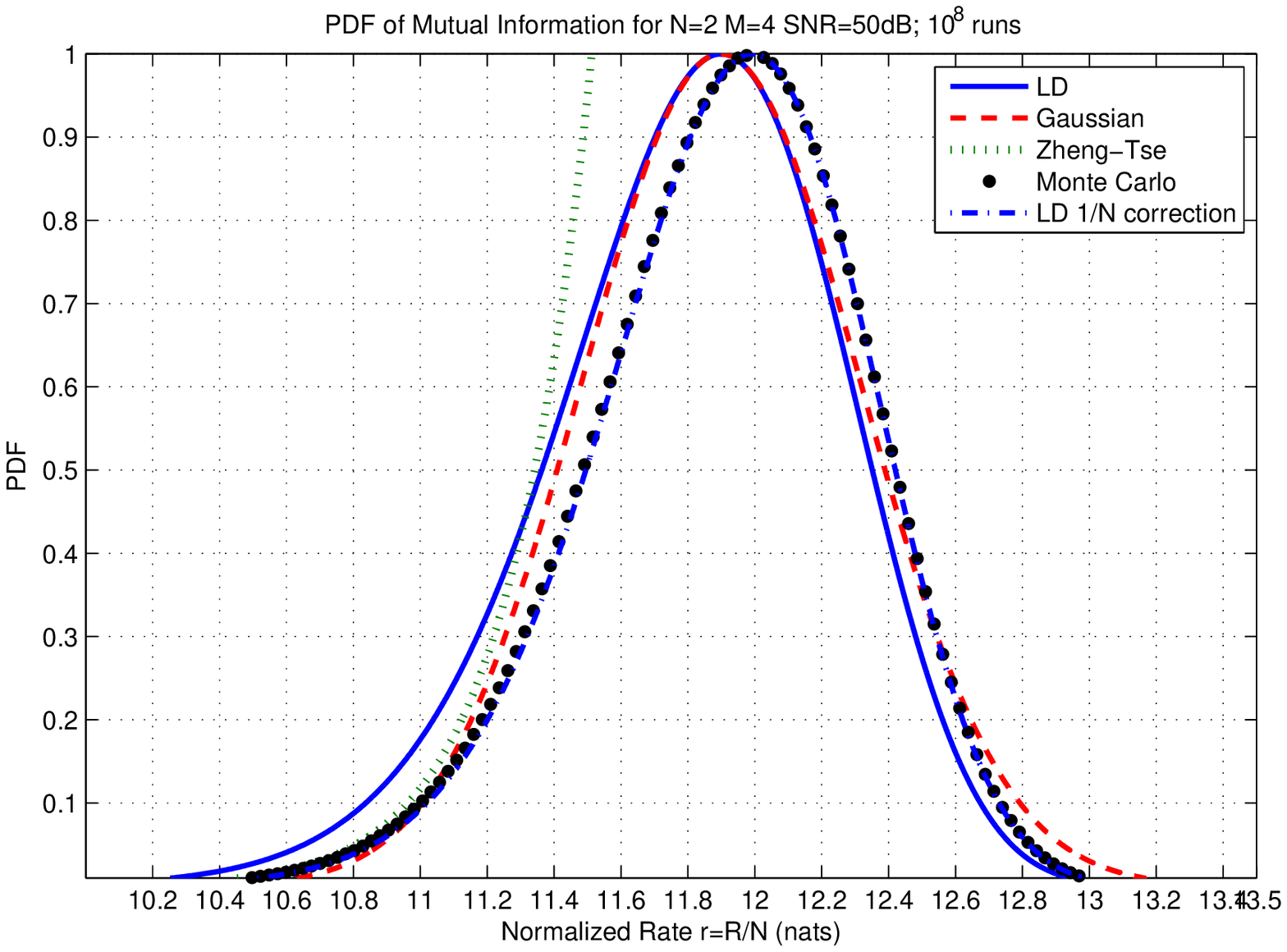}}}
\caption{Normalized probability distribution curves for the PDF of the mutual information for the antenna array $N=2$, $M=4$ for $\rho=20dB$ (a) and $\rho=50dB$ (b). In addition to the LD and Gaussian approximations and the Monte Carlo-generated curves, we have plotted the LD approximation including the $O(1/N)$ correction analyzed in Appendix \ref{app:1/Ncorrections}. We see that the latter curve agrees very well with the numerical one.} \label{fig:PDF_N2M4}
\end{figure*}

In Appendix \ref{app:1/Ncorrections} we provide an improved estimate on the probability distribution close its center. This estimate is a result of the inclusion of the $O(1/N)$ corrections to the distribution derived in \cite{Moustakas2003_MIMO1}. Fig. \ref{fig:PDF_N2M4} shows the normalized probability distribution function of the Gaussian approximation as well as the LD approximation with and without the $O(1/N)$ corrections. We see that the improved estimate behaves extremely well when the antenna numbers are quite small, in which cases the leading approximation (without the $O(1/N)$ correction), has some small discrepancies. (This should be contrasted with Fig. \ref{fig:E1_N5_beta2_SNR_100}, where $N=5$ and the $O(1/N)$ correction is no longer necessary to provide close agreement.)

\section{Conclusion}
\label{sec:conclusion}

In this paper we have used a large deviation approach, first introduced in the context of statistical mechanics \cite{Dyson1962_DysonGas, Vivo2007_LargeDeviationsWishart}, to calculate the probability distribution of the mutual information of MIMO channels in the limit of large antenna numbers. In contrast to previous approaches that focused only close to the mean of the distribution,\cite{Moustakas2003_MIMO1, Hochwald2002_MultiAntennaChannelHardening, Hachem2006_GaussianCapacityKroneckerProduct}, we also calculate the probability for {\it rare events} in the tails of the distribution, corresponding to instances where the observed mutual information differs by $\bigoh(N)$  from the most probable value of the asymptotic distribution (where the Gaussian approximation for the mutual information is invalid). We find that the distribution in those tails is markedly different from what happens near the mean and our resulting probability distribution interpolates seamlessly between the Gaussian approximation for rates close to the ergodic mutual information and the results of \cite{Zheng2003_DiversityMultiplexing} for large signal to noise ratios (where the outage probability is given asymptotically by (\ref{eq:DMT_def})). Our method thus provides an analytic tool to calculate outage probabilities at any point in the $(R, \rho, N)$ parameter space, as long as $N$ is large enough. We performed numerical simulations that showed the robustness of our approximation over a wide range of parameters.

Additionally, this approach also provides the probability distribution of eigenvalues constrained in the subset where the mutual information is fixed to $R$ for a given signal to noise ratio $\rho$. Interestingly, this eigenvalue density is of the form of the Mar\v{c}enko-Pastur distribution with square-root singularities. Since the outage probability is an increasing function of the rate $r$ for fixed $\rho$, we may use our approach to evaluate the transmission rate $R$ for a required outage $P_{out}$ and $\rho$. Thus, if the channel is known at the transmitter, we can optimize the transmitted rate by waterfilling on the known eigenvalue density that corresponds to the required outage probability \cite{Ordonez2009_OrderedEigsMIMO}. This generalization is left for a future work.

Finally, it is worth pointing out that, to our knowledge, this is the first time this methodology has been applied to information theory and communications, and it is our belief that it may find other applications in this field.  We can corroborate this belief by pointing out that this Coulomb gas methodology can be generalized to other channel distributions, as long as the resulting distribution can be written as a product of functions of the eigenvalues of $\bH^\dagger\bH$. Another related generalization is, for example, to include the correlations of the channel, a problem which is considerably more difficult compared to the present one. Some preliminary mathematical tools have already been developed in \cite{Matytsin1994_LargeNLimitIZIntegral}, and we will expand on this in the future.


\appendices

\section{Properties of tame probability measures}
\label{app:tameness}


This appendix is largely devoted to the study of the energy functional $\energy$:
\begin{align}
\tag{\ref{eq:energy}}
\energy[p]		&= \int \!x  p(x) \dd x- (\beta-1)\int\! p(x) \log x \dd x \\ \nonumber
			&-\iint p(x) p(y) \!\log|x-y| \dd x \dd y
\end{align}
where $p\in\tame$ is a tame density. As evidenced by definition \ref{dfn:tame} where the concept of tameness was introduced, an extremely important part in our analysis will be played by the so-called $L^{r}$ norm $\|\cdot\|_{r}$ defined by:
\begin{equation}
\|f\|_{r} \equiv \left(\int |f(x)|^{r} \dd x\right)^{1/r}.
\end{equation}
If a function $f$ has finite $L^{r}$ norm it is called $L^{r}$-integrable and the space of such functions constitutes a complete vector space (also denoted by $L^{r}$). The completeness of this space follows from {\it H\"older's inequality} which we state without proof and which will be of great use to us \cite{Fo99}:
\begin{equation}
\|fg\|_{1} \leq\|f\|_{r}\|g\|_{s}
\end{equation}
whenever the exponents $r,s>1$ are {\it conjugate}, that is: $r^{-1} + s^{-1} = 1$.

We will also make heavy use of the convolution $f*g$ between two functions $f$ and $g$:
\begin{equation}
(f*g)(x)	=\int f(x-y) g(y) \dd y.
\end{equation}
If $f\in L^{1}$ and $g\in L^{r}$, {\it Young's inequality} (pp. 240--241 in \cite{Fo99}) states that their convolution will be finite for almost every $x$ and also that:
\begin{equation}
\|f*g\|_{r} \leq\|f\|_{1}\|g\|_{r}.
\end{equation}

We may now proceed with the proof of lemma \ref{lem:tame} regarding the domain of $\energy$ and its continuity properties:
\begin{IEEEproof}[Proof of Lemma \ref{lem:tame}]
To show that $\energy$ is finite for all tame functions $p\in\tame$, we will study $\energy[p]$ term by term. To that end, let $p:\RR_{+}\to\RR$ be tame for some exponent $\eps>0$; that is, assume that $\int |p|^{1+\eps} <\infty$ and that $\int x p(x) \dd x<\infty$. We then have:
\begin{itemize}
\item The first term of $\energy[p]$ is finite by definition.
\item The second term in (\ref{eq:energy}) can be written as:
\begin{multline}
\left|\int p(x)\log x \dd x\right|	\leq \int |p(x) \log x|\dd x\notag\\
							= \int_0^1 |p(x) \log x| \dd x + \int_1^\infty |p(x) \log x| \dd x.
\end{multline}
Since $\log x < x$ for $x>1$, the second integral will be bounded from above by $\int x |p(x)| \dd x<\infty$. As for the first integral, set $r=1+\eps$ and $s=1+\frac{1}{\eps}$ so that $r^{-1}+s^{-1}=1$. Now, if $\chi_{[0,1]}$ is the indicator function of $[0,1]$, note that $\int |\chi_{[0,1]} \log x|^{s} \dd x = \int_{0}^{1}|\log x|^{s} \dd x<\infty$ for all $s>-1$. As a result, H\"older's inequality yields:
\begin{align}
\int_{0}^{1} |p(x) \log x| \dd x		&= \|p \cdot \chi_{[0,1]}\log\|_{1}\notag\\
							&\leq \|p\|_{1+\eps} \cdot \|\chi_{[0,1]}\log\|_{1+1/\eps}<\infty
\end{align}
on account of $p$ being $L^{1+\eps}$-integrable.
\item For the last term of $\energy$, let $D_{+} = \{(x,y)\in\RR^{2}: y>x\}$ and note that:
\begin{multline}
\left|\iint p(x) p(y) \log|x-y| \dd y\dd x\right|\\
	\shoveleft{\leq 2\iint_{D_{+}} |p(x) p(y) \log|x-y|| \dd y \dd x}\\
	= 2\int_{0}^{\infty} |p(x)|\int_{x}^{\infty} |p(y) \cdot \log(y-x)| \dd y \dd x.
\end{multline}
Now, the innermost integral can be written in the form:
\begin{multline}
\label{eq:estimate}
\int_{x}^{\infty} |p(y)|\cdot |\log(y-x)| \dd y\\
	\shoveleft =\int_{0}^{\infty} |p(x+w)\cdot\log w| \dd w\\
	\shoveleft =\int_{0}^{\infty} |p(x+w) K(w)| \dd w + \int_{1}^{\infty} |p(x+w)\log w| \dd w\\
	\leq \int_{0}^{\infty} |p(y)| K(y-x) \dd y \\
   + \int_{0}^{\infty} |p(1+x+w)\log (1+w)| \dd w.
\end{multline}
where $K(w)$ is the kernel:
\begin{equation}
K(w) = \begin{cases}\log|w|,	&0<w\leq 1\\
				0,		&\text{otherwise.}\end{cases}
\end{equation}
As above, $K$ will be $L^{s}$-integrable for all $s>-1$ and, in particular, for $s=1+\frac{1}{\eps}$. Therefore, we will have:
\begin{multline}
\int_{0}^{\infty} |p(x)| \int_{0}^{\infty} |p(y)| K(x-y) \dd y \dd x \\
= \big\||p|\cdot \big(|p|*|K|\big)\big\|_{1}\\
		\leq\|p\|_{1+\eps}\cdot \|p*K\|_{1+1/\eps}\\
		\leq \|p\|_{1+\eps} \cdot \|p\|_{1} \cdot \|K\|_{1+1/\eps}<\infty
\end{multline}
where the penultimate step is an application of H\"older's estimate and the last one follows from Young's inequality.

Finally, the second integral of (\ref{eq:estimate}) can be estimated by:
\begin{multline}
\label{eq:estimate2}
\int_{0}^{\infty} |p(1+x+w) \log(1+w)| \dd w\\
	\leq \int_{0}^{\infty} |p(1+x+w)| w \dd w\\
	\leq C x \int_{0}^{\infty} w |p(w)| \dd w
\end{multline}
for some sufficiently large $C>0$. Then, since $p$ is tame (i.e. $\int w |p(w)| \dd w <\infty$), we may integrate (\ref{eq:estimate2}) over $x$ to finally obtain that $\energy[p]<\infty$.
\end{itemize}

This completes the proof that $\energy[p]$ is finite for all tame functions $p\in\tame$. To show that $\energy$ is continuous on all subspaces of $L^{1+\eps}$-integrable functions with finite absolute mean, it simply suffices to note that all our estimates of $\energy[p]$ are bounded by the $L^{1+\eps}$ norm of $p$.
\end{IEEEproof}

\begin{remark*}
If a function is in $L^{r}$ for some $r>1$ and has finite mean, it will necessarily be in $L^{1}$ as well; in this way, tame measures form a (dense) subspace $\tame$ of $L^{1}(\RR_{+})$ that is similar to the union $\bigcup_{\eps>0}L^{1+\eps}$.
\end{remark*}
We will now prove Lemma \ref{lem:convex} showing that $\energy$ is not only continuous but also convex over the (convex) domain $\domain$ of tame {\it probability} measures.
\begin{IEEEproof}[Proof of Lemma \ref{lem:convex}]
Let $p,q\in\domain$ be two tame probability measures and introduce the bilinear pairing:
\begin{equation}
\langle p,q\rangle = -\int\int p(x) q(y) \log|x-y| \dd x \dd y
\end{equation}
which is actually well-defined on the whole space $\tame$ (as can be seen by the proof of lemma \ref{lem:tame}). Since the first two terms of $\energy$ are linear (and hence convex), it will suffice to show that:
\begin{equation}
\label{eq:convexpair}
\big\langle (1-t)p + tq, (1-t)p +tq\big\rangle < (1-t) \langle p, p\rangle + t \langle q, q \rangle
\end{equation}
for all $t\in(0,1)$. Indeed, if we let $\phi = p-q\in\tame$, equation (\ref{eq:convexpair}) reduces to showing that the pairing $\langle\cdot,\cdot\rangle$ is an inner product on the subspace of densities with zero total charge, i.e. that:
\begin{equation}
\langle \phi, \phi \rangle >0
\end{equation}
for any nonzero tame $\phi\in\tame$ with $\int \phi(x)\dd x=\int \big(p(x)-q(x)\big) \dd x=0$.

From the point of view of electrostatics, this is plain to see: after all $\langle \phi,\phi\rangle$ is just the self-energy of the charge density $\phi$. More specifically, let us define $D_{+}=\{(x,y): x<y\}$ as in the proof of lemma \ref{lem:tame}. Then we will have:
\begin{eqnarray}
\langle \phi,\phi\rangle	&=& -2\int_{D_{+}}\!\! \phi(x)\phi(y) \log|x-y| \dd x \dd y\nonumber\\
					&=& -2\int_{0}^{\infty}\phi(x) \int_{0}^{x} \phi(y) \log(x-y) \dd y \dd x \nonumber\\
					&>& 2\int_{0}^{\infty} \phi(x) \int_{0}^{x} \phi(y) (y-x) \dd y \dd x
\end{eqnarray}
So, if we set $\Phi (x) = \int_{0}^{x} \phi(y)\dd y$ and integrate by parts, we get:
\begin{eqnarray}
\langle\phi,\phi\rangle	&>& \int_{0}^{\infty}\phi(x)\int_{0}^{x} y\phi(y) \dd y \dd x - \int_{0}^{\infty} x\phi(x) \Phi(x) \dd x \nonumber\\
					&=& - \int_{0}^{\infty}\phi(x) \left(\int_{0}^{x}\Phi(y) \dd y\right)\dd x\nonumber\\
					&=& \int_{0}^{\infty}\Phi^{2}(x) \dd x-\Phi(\infty)\int_{0}^{\infty}\Phi(y) \dd y\nonumber\\
					&>& 0
\end{eqnarray}
since $\Phi(\infty)\equiv\int_{0}^{\infty}\phi(x)\dd x = 0 = \Phi(0)$ on account of $\phi$ having zero total charge.
\end{IEEEproof}

\section{Construction of the Coulomb gas model}
\label{app:Coulomb_gas}

In this appendix we will briefly show how the transition from discrete to continuous eigenvalue measures discussed in Section \ref{sec:map_coulomb_gas} occurs. As in the main text, we will not present any formal proof here either. However, we will argue that treating the formally discrete distribution of eigenvalues appearing in  (\ref{eq:E_lambda}), as continuous in the large $N$ limit is quite reasonable. A more formal method showing the same result appears in \cite{Vivo2007_LargeDeviationsWishart}. The main reasoning, also discussed in the main text, is that the external confining potentials defined by the first two terms in (\ref{eq:E_lambda}) or (\ref{eq:energy}) are strong enough to overcome the logarithmic repulsion between eigenvalues (third term in (\ref{eq:E_lambda})), and therefore guarantee that (with high probability) most of the eigenvalues will be confined in a finite width region near the minimum of the external potential. At the same time, this will mean that the eigenvalue density per unit length will be scaling with $N$ if $N$ is large enough. As a result, this can be seen as a  high-density limit and therefore the continuous approximation for the measure will be valid, at least close to configurations whose energy is low enough.

In the remainder of this section we will motivate the transition from the discrete to continuous eigenvalue densities and show what kind of terms we expect to see. We start by focusing in a finite region of eigenvalues of length $D$. We then divide the integration over $\lambda_k$ in (\ref{eq:CDF_vol_ratio}) in $L$ segments of length $\ell$, such that $L\ell=D$. The length of each segment $\ell$ has to be small enough so that the energy (\ref{eq:E_lambda}) can be well approximated with all eigenvalues within a given segment being placed at the endpoint of the segment. At the same time, it has to be large enough so that there is a macroscopic (i.e. $\bigoh(N)$) number of eigenvalues inside each segment. In principle, at the end of this exercise we need to take the limit $\ell\rightarrow 0$ as well, however we will discuss the subtleties of this limit later on. As a result, the integral over $\Dc \lambdav$ can be written as:
\begin{flalign}
\label{eq:entropy1}
  \int \Dc\lambdav	&\sim \prod_{k=1}^N \left(\sum_{m_k=1}^L \ell\right) = \prod_{m=1}^L \left(\sum_{n_m=0}^N \right)
  \frac{N!\ell^N }{\prod_{m=1}^L n_m!}
  \\
				&\sim  \prod_{m=1}^L \left(\sum_{n_m=0}^N\right)  \exp\left[-N\ell\sum_m p(m\ell) \log \left(p(m\ell)\right)\right]
  \label{eq:entropy2}
\end{flalign}
where $n_m$ are the number of $\lambda_k$'s that appear in the $m$th segment, with constraint $\sum_m n_m=N$. The factorials appearing at the RHS of (\ref{eq:entropy1}) are the number of ways the $N$ eigenvalues can be re-arranged in $L$ segments. This factor constitutes the entropy term and, for large $N$ and $n_m$, we can apply Stirling's formula to get the exponent in (\ref{eq:entropy2}) (where $p(m\ell) = n_m/(N\ell)$ is the fraction of eigenvalues per unit length appearing in segment $m$).

We next look at the form of the energy in (\ref{eq:E_lambda})
\begin{eqnarray}\label{eq:E_lambda_approx}
E(\lambdav) &\sim& \ell \sum_m p(m\ell) \left(m\ell - (\beta-1) \log m\ell\right) \\ \nonumber
&+& \ell^2 \sum_{m\neq m'} p(m\ell) p(m'\ell) \log\left|(m-m')\ell\right| \\ \nonumber
&+& \frac{\ell}{N} \sum_m p(m\ell) \log a_m \ell
\end{eqnarray}
The last term captures the repulsive interaction between eigenvalues in the same segment $m$. The value of $a_m$ represents the typical distance between eigenvalues in segment $m$ in units of $\ell$ and therefore is a number of order unity. We may now let $\ell\rightarrow 0$, which will make the sums converge to integrals $\ell \sum_m \rightarrow \int dx$ and $p(m\ell)$ can be written as a continuous function $p(x)$. Representing the sum over all possible states (i.e. the product of sums in (\ref{eq:entropy2})) by $\int \Dc p$ we can now get
\begin{equation}\label{eq:partition function}
    \pf \sim \int_\chi \!\Dc p\, e^{-N^2\energy[p]} e^{-N\int dx p(x)\log p(x)} e^{N\int dx p(x) \log d(x)}
\end{equation}
where $d(x)=a_m \ell$ is the average distance between eigenvalues at the position $x=m\ell$. One can estimate this average inter-eigenvalue distance to be
\begin{equation}\label{eq:inter_eig_d}
    d(x) \sim a_m\ell \sim \frac{1}{Np(x)}
\end{equation}
This was first proposed by Dyson \cite{Dyson1962_DysonGas, Vivo2007_LargeDeviationsWishart, Mehta_book} and was shown explicitly more recently in \cite{Brezin1993_UniversalEigCorrsRMT}. It is remarkable that with this choice of $d(x)$ the $\bigoh(N)$ dependence on $p(x)$ in the exponent of (\ref{eq:partition function}) vanishes. This surprising fact is true only for complex matrices \cite{Mehta_book} in which, up to uninteresting constants, the leading correction to the $N^2\energy[p]$ term in the exponent is $\bigoh(1)$.

\section{Solution of the Variational Equation}
\label{app:solution}

In this appendix, we give a more detailed account of the solution of the variational equation:
\begin{equation*}
\delta\lagrange_{1}[p]=0
\end{equation*}
where $\lagrange_{1}$ is the Lagrangian function of (\ref{eq:L1}). To that end, if $\phi\in\tame$ is tame, we get:
\begin{flalign}
\label{eq:vardiff}
\lagrange_{1}[p+t\phi]	&= \lagrange_{1}[p] + t\lagrange_{1}[\phi]\notag\\
					&- 2t \iint \phi (x) p(y) \log|x-y| \dd y \dd x + \bigoh(t^{2})
\end{flalign}
and a simple differentiation at $t=0$ yields:
\begin{multline}
\langle\delta \lagrange_{1}[p],\phi\rangle	=\left.\frac{d}{dt}\right|_{t=0}\!\!\!\lagrange_{1}[p+t\phi]\\
			= \lagrange_{1}[\phi] - 2\iint \phi(x) p(y) \log|x-y|\dd y \dd x\\
			=\int \phi(x) \Psi[p,x]\dd x,
\end{multline}
where the expression $\Psi[p,x]$ is given by:
\begin{flalign}
\Psi[p,x]	&= 2\!\!\int p(y) \log|x-y|\dd y - x \notag\\
			&+(\beta -1)\log x+c+k\log(1+\rho x) + \nu(x).
\end{flalign}
Thus, for the above expression to vanish identically for all $\phi\in\tame$, we must have $\Psi[p,x] = 0$, and this is precisely (\ref{eq:func_deriv_E1_result}), repaeted below:
\begin{multline}
  2\int_0^\infty p(x')\log|x-x'| \dd x'	=  x - (\beta-1)\log x \\
  								- c - k\log(1+\rho x)-\nu(x).
  \end{multline}

Having derived this stationarity equation in terms of $p$, we will devote the rest of this appendix to the expression (\ref{eq:func_deriv_E1_result_3}), also repeated below for convenience, that is obtained after differentiating (\ref{eq:func_deriv_E1_result}) above:
\begin{equation}
2{\cal P} \int_a^b \frac{p(y)}{x-y} dy = 1-\frac{\beta-1}{x}-\frac{k\rho}{1+\rho x} \equiv f(x)
\end{equation}
for all $x\in[a,b]$ (cf. section \ref{sec:integral_equation}). This integral equation is known as the {\it airfoil equation} and can be studied with the help of the {\it finite} Hilbert transform \cite{Tricomi_book_IntegralEquations}:
\begin{equation}
\hilbert[\phi](x) = \mathcal{P}\!\int_{-1}^{1} \frac{\phi(y)}{y-x} \dd y.
\end{equation}

If $r>1$, the $\hilbert$-transform maps $L^{r}$ to $L^{r}$ but, nevertheless, it lacks a unique inverse.\footnote{This is a remarkable difference from the case of the {\it infinite} Hilbert transform which integrates over all $\RR$ and which {\it is} invertible \cite{Tricomi_book_IntegralEquations}.} Indeed, the kernel of $\hilbert$ is spanned by the function $\omega(x) = (1-x^{2})^{-\frac{1}{2}}$: $\hilbert[\omega](x) = 0$ for all $x\in(-1,1)$. Outside this kernel, the solutions $\phi$ to the airfoil equation $\hilbert[\phi] = g$ with $\phi,g\in L^{r}[-1,1]$ will satisfy \cite{Tricomi_book_IntegralEquations}:
\begin{equation}
\label{eq:inverseHilbert}
\phi(x) = -\frac{1}{\pi}\mathcal{P} \int_{-1}^{1} \sqrt{\frac{1-y^{2}}{1-x^{2}}} \frac{g(y)}{y-x} \dd y + \frac{c}{\sqrt{1-x^{2}}}
\end{equation}
where $c$ is an arbitrary constant that stems from the fact that any two solutions of the airfoil equation differ by a multiple of $\omega(x) = (1-x^{2})^{-\frac{1}{2}}$.

Hence, after rescaling the interval $[-1,1]$ to $[a,b]$, the solution of the stationarity equation (\ref{eq:func_deriv_E1_result_3}) will be given by:
\begin{equation}
p(x)	= \frac{{\cal P} \int_a^b  \frac{\sqrt{(y-a)(b-y)}f(y)}{y-x} dy + C'}{2\pi^2\sqrt{(x-a)(b-x)}}
\end{equation}
whenever $f$ is itself $L^{1+\eps}$-integrable. So, by substituting $f(x) = 1- \frac{\beta-1}{x} - \frac{k}{x+z}$ from (\ref{eq:func_deriv_E1_result_3}) and performing one last integration, we obtain the final result (\ref{eq:gen_solution_int_eq0}).

It is worthwhile to mention here again how this procedure breaks down if we allow the support of $p$ to extend to $a=0$ for $\beta>1$: in that case, the function $f$ also extends all the way to $a=0$ and the term $\frac{\beta-1}{x}$ makes it non-integrable. However, since the Hilbert transform preserves $L^{r}$-integrability for $r>1$ and $p$ is assumed tame (and hence $L^{1+\eps}$-integrable), equation (\ref{eq:func_deriv_E1_result_3}) would equate an integrable function with a non-integrable one, thus yielding a contradiction. Therefore, as we stated in section \ref{sec:integral_equation}, solutions with $a=0$ are physically inadmissible when $\beta>1$.

\section{Proof of uniqueness of solution of (\ref{eq:p_a_=0}),(\ref{eq:norm_integral})}
\label{app:uniqueness_ab}

In order to show that (\ref{eq:p_a_=0}), (\ref{eq:norm_integral}) admit a unique
solution, we start by observing that for fixed $k,\beta$ and $z$, (\ref{eq:p_a_=0}) has a unique positive solution $a\leq b$. Then, from the implicit function theorem, this solution can be captured in terms of $b$ by a smooth function $a(b)$ whose derivative can be obtained implicitly from (\ref{eq:p_a_=0}) (and which is negative). With this in mind, the normalization integral $g(b) = \int_{a(b)}^{b}p(x) \dd x$ takes the form:
\begin{equation}
g(b) = \frac{a(b)+b}{4} + \frac{1}{2}\left(\rho^{-1}-k-(\beta-1)\left(1+\frac{1}{\rho\sqrt{a(b)b}}\right)\right)\notag
\end{equation}
and this is actually an increasing function of $b$. Indeed, after a somewhat painful calculation, one obtains:
\begin{equation}
g'(b) = \frac{\rho}{4}\left[1+\frac{(\beta-1)}{\rho\sqrt{a(b)\,b^{3}}}\right]\frac{b-a(b)}{1+\rho b}>0
\end{equation}
However, with $a(b)$ decreasing and bounded below by $0$, this last equation yields $g'(b)>1/8$ for large enough $b$, i.e. $\lim_{b\to\infty} g(b) = +\infty$. So, by continuity, there will be a (necessarily) unique $b^{*}$ such that $g(b^{*})=1$. Hence, the pair $a^{*}=a(b^{*}), b=b^{*}$ will be the unique solution to (\ref{eq:p_a_=0}), (\ref{eq:norm_integral}).


\section{$O(1/N)$ correction to the LD approximation}
\label{app:1/Ncorrections}

Here we provide an improved estimate on the probability distribution close to the center of the distribution. This estimate is a result of the inclusion the $O(1/N)$ higher moment corrections to the distribution derived in \cite{Moustakas2003_MIMO1}.

It is well known\cite{Bouchaud_book_FinancialRiskDerivativePricing} that to provide asymptotic corrections to the limiting Gaussian distribution due to the presence of a small (but finite) skewness we need to change the distribution as follows:
\begin{equation}\label{eq:correction_to CLT}
    P_N(x)=\frac{e^{-\frac{x^2}{2v}}}{\sqrt{2\pi v}}\left(1-\frac{s}{2v^2}\left(x-\frac{x^3}{3 v}\right)\right)
\end{equation}
where $v$ is the variance of the asymptotically Gaussian distribution and $s$ is the third moment of the distribution. Clearly, the above distribution cannot be valid over the entire support of $x$ since the cubic polynomial will become negative for some value of $x$. Nevertheless, since the third moment is small for large $N$ this value of $x$ will become asymptotically large.

We may therefore apply the above formula to our model. The value of the third moment $s=s_3/N$ has been calculated in [(60) in \cite{Moustakas2003_MIMO1}] and it is of order $O(1/N)$. As a result, the correction to the Gaussian approximation of the mutual information is given by
\begin{eqnarray}\label{eq:correction_to Gaussian}
    P_N(R) &=& \frac{e^{-\frac{(R-Nr_{erg})^2}{2v_{erg} } } }{\sqrt{2\pi v_{erg}}} \cdot \\ \nonumber
    && \left(1-\frac{s_3}{2Nv_{erg}^2}(R-Nr_{erg}) + \frac{(R-Nr_{erg})^3}{3 v_{erg}}\right)
\end{eqnarray}
To order $O(1/N)$, there is also the correction to the mean of the mutual information \cite{Moustakas2003_MIMO1}, which needs to be subtracted off from $I_N$.

Now, to obtain the correction to the LD approximation, we need to take into account that the large deviations function $\Ec_1$ also has a cubic term for $r\approx r_{erg}$, which needs to be balanced. This can be done by adding a cubic term that cancels this term for $r\approx r_{erg}$. Thus we obtain
\begin{eqnarray}\label{eq:correction_to LD}
    P_N(r)&=&\frac{Ne^{-N^2 (\Ec_1(r)-\Ec_0)}}{\sqrt{2\pi v_{erg}}} \left(1-\frac{s_3}{2v_{erg}^2}(r-r_{erg}) \right. \nonumber \\ &+&\left.\frac{N^2}{6}\left(\frac{s_3}{v_{erg}^3}+s_{erg}\right)(r-r_{erg})^3  \right)
\end{eqnarray}
where $s_{erg}$ is given by (\ref{eq:E3_asymptotics}).


\begin{thebibliography}{10}
\providecommand{\url}[1]{#1}
\csname url@samestyle\endcsname
\providecommand{\newblock}{\relax}
\providecommand{\bibinfo}[2]{#2}
\providecommand{\BIBentrySTDinterwordspacing}{\spaceskip=0pt\relax}
\providecommand{\BIBentryALTinterwordstretchfactor}{4}
\providecommand{\BIBentryALTinterwordspacing}{\spaceskip=\fontdimen2\font plus
\BIBentryALTinterwordstretchfactor\fontdimen3\font minus
  \fontdimen4\font\relax}
\providecommand{\BIBforeignlanguage}[2]{{%
\expandafter\ifx\csname l@#1\endcsname\relax
\typeout{** WARNING: IEEEtran.bst: No hyphenation pattern has been}%
\typeout{** loaded for the language `#1'. Using the pattern for}%
\typeout{** the default language instead.}%
\else
\language=\csname l@#1\endcsname
\fi
#2}}
\providecommand{\BIBdecl}{\relax}
\BIBdecl
\renewcommand{\BIBentryALTinterwordstretchfactor}{4}

\bibitem{Zheng2003_DiversityMultiplexing}
L.~Zheng and D.~N.~C. Tse, ``Diversity and multiplexing: {A} fundamental
  tradeoff in multiple-antenna channels,'' \emph{{IEEE} Trans. Inform. Theory},
  vol.~49, no.~5, pp. 1073--1096, May 2003.

\bibitem{Foschini1998_BLAST1}
G.~J. Foschini and M.~J. Gans, ``On limits of wireless communications in a
  fading environment when using multiple antennas,'' \emph{Wireless Personal
  Communications}, vol.~6, pp. 311--335, 1998.

\bibitem{Telatar1995_BLAST1}
I.~E. Telatar, ``Capacity of multi-antenna {G}aussian channels,''
  \emph{European Transactions on Telecommunications and Related Technologies},
  vol.~10, no.~6, pp. 585--596, Nov. 1999.

\bibitem{Rapajic2000_InfoCapacityOfARandomSignatureMIMOChannel}
P.~B. Rapajic and D.~Popescu, ``Information capacity of a random signature
  multiple-input multiple-output chanel,'' \emph{{IEEE} Trans. Commun.},
  vol.~48, no.~8, p. 1245, Aug. 2000.

\bibitem{Wang2002_OutageMutualInfoOfSTMIMOChannels}
Z.~Wang and G.~B. Giannakis, ``Outage mutual information of space-time {MIMO}
  channels,'' \emph{{IEEE} Trans. Inform. Theory}, vol.~50, no.~4, pp.
  657--662, Apr. 2004.

\bibitem{Biglieri1998_FadingChannels}
E.~Biglieri, J.~Proakis, and S.~Shamai, ``Fading channels:
  Information-theoretic and communications aspects,'' \emph{{IEEE} Trans.
  Inform. Theory}, vol.~44, no.~6, p. 2619, Oct. 1998.

\bibitem{Moustakas2003_MIMO1}
A.~L. Moustakas, S.~H. Simon, and A.~M. Sengupta, ``\mbox{MIMO} capacity
  through correlated channels in the presence of correlated interferers and
  noise: \mbox{A} (not so) large \mbox{N} analysis,'' \emph{{IEEE} Trans.
  Inform. Theory}, vol.~49, no.~10, pp. 2545--2561, Oct. 2003.

\bibitem{Hochwald2002_MultiAntennaChannelHardening}
B.~M. Hochwald, T.~L. Marzetta, and V.~Tarokh, ``Multi-antenna channel
  hardening and its implications for rate feedback and scheduling,''
  \emph{{IEEE} Trans. Inform. Theory}, vol.~50, no.~9, pp. 1893--1909, Sept.
  2004.

\bibitem{Hachem2006_GaussianCapacityKroneckerProduct}
W.~Hachem, O.~Khorunzhiy, P.~Loubaton, J.~Najim, and L.~Pastur, ``A new
  approach for capacity analysis of large dimensional multi-antenna channels,''
  \emph{{IEEE} Trans. Inform. Theory}, vol.~54, pp. 3987--4004, Sep. 2008.

\bibitem{Taricco2006_MIMOCorrelatedCapacity}
G.~Taricco, ``On the capacity of separately-correlated {MIMO} {R}ician fading
  channels,'' \emph{Proc. IEEE Globecom 2006}, Dec. 2006.

\bibitem{Taricco2008_MIMOCorrelatedCapacity}
------, ``Asymptotic mutual information statistics of separately-correlated
  {MIMO} {R}ician fading channels,'' \emph{{IEEE} Trans. Inform. Theory},
  vol.~54, no.~8, p. 3490, Aug. 2008.

\bibitem{Azarian2007_finite_rate_DMT}
K.~Azarian and H.~El-Gamal, ``The {T}hroughput {R}eliability {T}radeoff in
  block-fading {MIMO} channels,'' \emph{{IEEE} Trans. Inform. Theory}, vol.~53,
  no.~2, p. 488, Feb. 2007.

\bibitem{Dyson1962_DysonGas}
F.~Dyson, ``Statistical theory of the energy levels of complex systems. {I},''
  \emph{J. Math. Phys.}, vol.~3, p. 140, 1962.

\bibitem{Majumdar2006_LesHouches}
S.~N. Majumdar, \emph{Random Matrices, the {U}lam Problem, Directed Polymers \&
  Growth Models, and Sequence Matching}, ser. Les Houches, M.~M{\'e}zard and
  J.~P. Bouchaud, Eds.\hskip 1em plus 0.5em minus 0.4em\relax Elsevier, July
  2006, vol. Complex Systems.

\bibitem{Vivo2007_LargeDeviationsWishart}
P.~Vivo, S.~N. Majumdar, and O.~Bohigas, ``Large deviations of the maximum
  eigenvalue in {W}ishart random matrices,'' \emph{J. Phys. A}, vol.~40, pp.
  4317--4337, 2007.

\bibitem{Vivo2008_DistributionsConductanceShotNoise}
------, ``Distributions of conductance and shot noise and associated phase
  transitions,'' \emph{Phys. Rev. Lett.}, vol. 101, p. 216809, 2008.

\bibitem{Nadal2009_NonIntersectingBrowianInterfaces}
C.~Nadal and S.~N. Majumdar, ``Nonintersecting brownian interfaces and wishart
  random matrices,'' \emph{Phys. Rev. E}, vol.~79, p. 061117, 2009.

\bibitem{Johansson1998_2ndOrderRMTFluctuations}
K.~Johansson, ``On fluctuations of eigenvalues of random hermitian matrices,''
  \emph{Duke Math. J.}, vol.~91, no.~1, pp. 151--204, 1998.

\bibitem{Tulino2004_RMTInfoTheoryReview}
A.~M. Tulino and S.~Verd{\'u}, ``Random matrix theory and wireless
  communications,'' \emph{Foundations and Trends in Communications and
  Information Theory}, vol.~1, no.~1, pp. 1--182, 2004.

\bibitem{Dean2008_ExtremeValueStatisticsEigsGaussianRMT}
D.~S. Dean and S.~N. Majumdar, ``Extreme value statistics of eigenvalues of
  {G}aussian random matrices,'' \emph{Phys. Rev E}, vol.~77, p. 041108, 2008.

\bibitem{Papoulis_book}
A.~Papoulis, \emph{Probability, Random Variables, and Stochastic Processes},
  3rd~ed.\hskip 1em plus 0.5em minus 0.4em\relax Singapore: McGraw-Hill, 1991.

\bibitem{Simon2002_TIMO1}
S.~H. Simon and A.~L. Moustakas, ``Optimizing {MIMO} systems with channel
  covariance feedback,'' \emph{{IEEE} J. Select. Areas Commun.}, vol.~21,
  no.~3, Apr. 2003.

\bibitem{Mehta_book}
M.~L. Mehta, \emph{Random Matrices}, 2nd~ed.\hskip 1em plus 0.5em minus
  0.4em\relax San Diego, CA: Academic Press, 1991.

\bibitem{Feynman1965_QM_PathIntegrals}
R.~P. Feynman and A.~R. Hibbs, \emph{Quantum Mechanics and Path
  Integrals}.\hskip 1em plus 0.5em minus 0.4em\relax New York: McGraw-Hill,
  1965.

\bibitem{Dembo_book_LargeDeviationsTechniques}
A.~Dembo and O.~Zeitouni, \emph{Large Deviations Techniques and
  Applications}.\hskip 1em plus 0.5em minus 0.4em\relax New York, USA:
  Springer-Verlag Inc., 1998.

\bibitem{Boyd_book}
S.~Boyd and L.~Vandenberghe, \emph{Convex Optimization}.\hskip 1em plus 0.5em
  minus 0.4em\relax Cambridge Univ. Press, 2004.

\bibitem{Tricomi_book_IntegralEquations}
F.~G. Tricomi, \emph{Integral Equations}, ser. Pure Appl. Math V.\hskip 1em
  plus 0.5em minus 0.4em\relax London: Interscience, 1957.

\bibitem{Mikhlin_book_IntegralEquations}
S.~G. Mikhlin, \emph{Integral Equations}.\hskip 1em plus 0.5em minus
  0.4em\relax New York: Pergamon, 1964.

\bibitem{Chen1996_EigDistributionsLaguerre}
Y.~Chen and S.~M. Manning, ``Some eigenvalue distribution functions of the
  laguerre ensemble,'' \emph{J. Phys. A: Math. Gen.}, vol.~29, pp. 7561--7579,
  1996.

\bibitem{Bender_Orszag_book}
C.~M. Bender and S.~A. Orszag, \emph{Advanced Mathematical Methods for
  Scientists and Engineers}.\hskip 1em plus 0.5em minus 0.4em\relax New York,
  NY: McGraw-Hill, 1978.

\bibitem{Verdu1999_MIMO1}
S.~Verd{\'u} and S.~Shamai, ``Spectral efficiency of \mbox{CDMA} with random
  spreading,'' \emph{{IEEE} Trans. Inform. Theory}, vol.~45, no.~2, pp.
  622--640, Mar. 1999.

\bibitem{Bender2000_HDR_ComMagReview}
P.~Bender, P.~Black, M.~Glob, R.~Padovani, N.~Sindhushayaba, and A.~Viterbi,
  ``\mbox{CDMA/HDR}: A bandwidth-efficient high-speed wireless data service for
  nomadic users,'' \emph{IEEE Communications Magazine}, pp. 70--77, Jul. 2000.

\bibitem{Ordonez2009_OrderedEigsMIMO}
L.~G. Ord\'{o}nez, D.~P. Palomar, and J.~R. Fonollosa, ``Ordered eigenvalues of
  a general class of hermitian random matrices with application to the
  performance analysis of mimo systems,'' \emph{{IEEE} Trans. Signal Process.},
  vol.~57, no.~2, pp. 672--689, 2009.

\bibitem{Matytsin1994_LargeNLimitIZIntegral}
A.~Matytsin, ``On the large-{N} limit of the {I}tzykson-{Z}uber integral,''
  \emph{Nuclear Physics B411}, pp. 805--820, 1994.

\bibitem{Fo99}
G.~B. Folland, \emph{Real Analysis}, 2nd~ed.\hskip 1em plus 0.5em minus
  0.4em\relax Wiley-Interscience, 1999.

\bibitem{Brezin1993_UniversalEigCorrsRMT}
E.~Br\'{e}zin and A.~Zee, ``Universality of the correlations between
  eigenvalues of large random matrices,'' \emph{Nuclear Physics B (FS)}, vol.
  402, pp. 613--627, 1993.

\bibitem{Bouchaud_book_FinancialRiskDerivativePricing}
J.-P. Bouchaud and M.~Potters, \emph{Theory of Financial Risk and Derivative
  Pricing}, 2nd~ed.\hskip 1em plus 0.5em minus 0.4em\relax Cambridge, UK:
  Cambridge, 2003.

\end{thebibliography}



\end{document}